\documentclass{article}
\usepackage{epsfig}
\usepackage{amssymb}
\usepackage{amsmath}
\usepackage{amsbsy}
\usepackage{wrapfig}
\usepackage{color}
\usepackage{graphicx}
\usepackage[colorlinks=true, pdfstartview=FitV, linkcolor=blue, citecolor=blue, urlcolor=blue]{hyperref}

\def\green#1{\textcolor[rgb]{0.1, 0.8,  0.3} {#1}}

\usepackage{verbatim}
\def\wt{\widetilde}

\def\wh{\widehat}

\def\ds{\displaystyle}

\renewcommand{\theequation}{\arabic{section}-\arabic{equation}}
\makeatletter
\@addtoreset{equation}{section}
\makeatother
\def\tr{\mathrm {Tr}}
\def\le{\left}
\def\ri{\right}

\def\Pf {\mathrm {Pf\ }}

\def\bc{\begin{corollary}}
\def\ec{\end{corollary}}
\def\a{\alpha}
\def\&{&{\hskip -20pt}}

\def \p{\mathbf p}
\def\q{\mathbf q}

\def\br{\begin{remark}\rm\small}
\def\1{{\bf 1}}
\def\er{\end{remark}}
\def\bt{\begin{theorem}}
\def\et{\end{theorem}}

\def\bx{\begin{examp}}
\def\ex{\end{examp}}
\def\bd{\begin{definition}}
\def\ed{\end{definition}}
\def\bp{\begin{proposition}\rm}
\def\bl{\begin{lemma}\em}
\def\el{\end{lemma}}
\def\ep{\end{proposition}}
\def\bea{\begin{eqnarray}}
\def\eea{\end{eqnarray}}

\def\C{{\mathbb C}}
\def\R{{\mathbb R}}
\def\N{{\mathbb N}}
\def\Z{{\mathbb Z}}

\newtheorem{assumption}{Assumption}[section]
\newtheorem{theorem}{Theorem}[section]
\newtheorem{examp}{Example}[section]
\newtheorem{coroll}{Corollary}[section]
\newtheorem{examps}{Examples}[section]

\newtheorem{lemma}{Lemma}[section]
\newtheorem{remark}{Remark}[section]
\newtheorem{remarks}[remark]{Remarks}
\newtheorem{proposition}{Proposition}[section] 
\newtheorem{definition}{Definition}[section]
\def\br{\begin{remark}}
\def\er{\end{remark}}
\def\bt{\begin{theorem}}
\def\et{\end{theorem}}
\def\bc{\begin{coroll}}
\def\ec{\end{coroll}}
\def\brs{\begin{remarks} \rm\
\begin{enumerate}}
\def\ers{\end{enumerate}\end{remarks}}
\def\bl{\begin{lemma}}
\def\el{\end{lemma}}
\def\bxs{\begin{examps}. \rm\begin{enumerate}}
\def\exs{\end{enumerate}\end{examps}}
\def\bd{\begin{definition}}
\def\ed{\end{definition}}
\def\bp{\begin{proposition}}
\def\ep{\end{proposition}}
\def\be{\begin{equation}}
\def\ee{\end{equation}}

\def\d{{\rm d}}
\def\bea{\begin{eqnarray}}
\def\eea{\end{eqnarray}}
\def\beas{\begin{eqnarray*}}
\def\eeas{\end{eqnarray*}}

\def\iint{\int\!\!\!\!\int}
\def\C{{\mathbb C}}

\def\R{{\mathbb R}}
\def\N{{\mathbb N}}

\def\Z{{\mathbb Z}}

\date{}
\begin{document}
%

\baselineskip 16pt plus 1pt minus 1pt
\vspace{0.2cm}
\begin{center}
\begin{Large}
\textbf{ The Cauchy two-matrix model}
\end{Large}\\
\bigskip
\begin{large} {M.
Bertola $^{\dagger\ddagger}$ \footnote{Work supported in part by the Natural
    Sciences and Engineering Research Council of Canada (NSERC)},  M. Gekhtman~$^a$ 
\footnote{Work supported in part by NSF Grant DMD-0400484.}, J. Szmigielski~$^b$ \footnote{Work supported in part by the Natural
    Sciences and Engineering Research Council of Canada (NSERC),
    Grant. No. 138591-04}}
\end{large}
\\
\bigskip
\begin{small}
$^{\dagger}$ {\em Centre de recherches math\'ematiques,
Universit\'e de Montr\'eal\\ C.~P.~6128, succ. centre ville, Montr\'eal,
Qu\'ebec, Canada H3C 3J7 \\
~~E-mail: bertola@crm.umontreal.ca}
\smallskip

$^{\ddagger}$ {\em Department of Mathematics and
Statistics, Concordia University\\ 1455 de Maisonneuve W., Montr\'eal, Qu\'ebec,
Canada H3G 1M8} \\
\smallskip
~$^a$ {\em Department of Mathematics
255 Hurley Hall, Notre Dame, IN 46556-4618, USA\\
~~E-mail: Michael.Gekhtman.1@nd.edu}

\smallskip
~$^b$ {\em Department of Mathematics and Statistics, University of Saskatchewan\\ 106 Wiggins Road, Saskatoon, Saskatchewan, S7N 5E6, Canada\\
~~E-mail: szmigiel@math.usask.ca}
\end{small}
\end{center}

\bigskip
\begin{center}{\bf Abstract}\\
\end{center}
We introduce a new class of two(multi)-matrix models of positive Hermitean matrices coupled in a chain; the coupling is related to the Cauchy kernel and differs from the exponential coupling more commonly used in similar models. The correlation functions are expressed entirely  in terms of certain biorthogonal polynomials and solutions of appropriate Riemann--Hilbert problems, thus  paving the way to a steepest descent analysis and universality results. The interpretation of the formal expansion of the partition function in terms of multicolored ribbon-graphs is provided and a connection to the $O(1)$ model. 
A steepest descent analysis of the partition function reveals that the model is related to a trigonal curve (three-sheeted covering of the plane) much in the same way as the Hermitean matrix model is related to a hyperelliptic curve.

\tableofcontents
\section{Introduction}
 In the last two decades or so, the theory of matrix models  have been
 an incredibly fertile ground for a fruitful interaction between  theoretical physics, statistics, analysis, number theory and dynamical systems (see the classical  \cite{MehtaBook} and references therein).

The interplay with analysis  has been particularly  beneficial for the Hermitean matrix model \cite{Deift}, largely due to the realization that the matrix model could be ``solved'' in terms of orthogonal polynomials. This allowed to translate questions about  the spectrum of large random matrices into questions about the asymptotics of orthogonal polynomials, connecting the former with a very well developed area of analysis. The availability of Riemann--Hilbert methods \cite{FIK0, DKMVZ} was the crucial ingredient in addressing questions of universality in the bulk and at the edge of the spectrum.

It should be mentioned that this fortunate symbiosis relies on the following features: 
\begin{itemize}
\item the possibility of rewriting the matrix integral in terms of eigenvalues and the correlation functions in terms of suitable orthogonal polynomials;
\item the Riemann--Hilbert characterization of orthogonal polynomials \cite{FIK0};
\item the Christoffel--Darboux formula, allowing to rewrite the kernel of the correlation functions in terms of only two orthogonal polynomials of degree $N$;
\item the nonlinear steepest descent method \cite{DKMVZ} applied to the RH problem for orthogonal polynomials.
\end{itemize}

There are by now several matrix models based on  various ensembles of matrices. For some (Symplectic, Orthogonal), a connection with  (skew-)orthogonal  polynomials can be established \cite{MehtaBook,WidomEnsembles} and some of the steps in the above list have been performed, but typically only for a certain restricted class of ``potentials''. For others (like $O(n)$ models, rectangular models) the methods used rely on symmetries of the integral and dynamical system approaches (the ``loop equations'') \cite{EynardOn, EynardLoop0, EynardLoop1, EynardOrantinLoop, AkemannLoops}

There are also the so--called {\em multi-matrix-models} which involve ensembles of several matrices (typically Hermitean) and one of the most studied among them  is the {\em Itzykson--Zuber--Harish-Chandra} (IZHC) chain of matrices. In its simplest form, the two-matrix model, it amounts  to the study of the spectral properties of a  pair of Hermitean matrices of size $N\times N$ with a  measure 
\be
\d \mu(M_1,M_2)= \d M_1 \d M_2 {\rm e}^{-N \tr (V_1(M_1)+V_2(M_2) - M_1M_2)}
\ee
 where the interaction term is ${\rm e}^{N\tr M_1M_2}$. The first three items of the bulleted list above can be implemented at least for potentials whose derivative is a rational function \cite{EynardMehta, BEH_dualityCMP, Berto_JAT1, Bertosemiclass} while the last item is still not fully under control. One of the main reasons for this difficulty is  that the size of the RHP {\em depends on the potentials}: for example if $V_j$ are polynomials of degree $d_j$ then there are two relevant RHPs of the same size. 
 Clearly the question arises as to whether a Riemann--Hilbert method can be utilized for a more general class of potentials, for example real-analytic, like in the case of the single-matrix model.
 
 The present paper is a part of a larger  project initiated in \cite{Paper1},  which puts forward a new multimatrix model (at the moment primarily two matrices) that can be completely solved along the lines described above.
 
While the new model is very close in spirit to the IZHC model, the different interaction 
\be
I(M_1,M_2) = \frac 1{\det(M_1+M_2)^N}
\ee
links the model to a new class of biorthogonal polynomials which were studied in 
\cite{Paper1} and termed  ``Cauchy biorthogonal polynomials''.
These polynomials  have several of the desirable features of classical orthogonal polynomials.

We will fill in more details in Sect. \ref{sect3}   but  here we just indicate that 
the model we want to study is defined on the space of pairs of {\em positive} Hermitean matrices of size $N$ equipped with a measure
\be
\d \mu(M_1,M_2) := \d M_1 \d M_2 \frac {\alpha(M_1) \beta(M_2)}{\det(M_1+M_2)^N} 
\ee
for {\bf arbitrary} measures $\alpha(x)\d x, \beta(y)\d y$ on $\R_+$ (to be understood  in the formula above as {conjugation-invariant measures on positive definite Hermitean matrices or, equivalently,} measures on the spectra of $M_i$'s). This immediately puts the problem at the same level of  generality as  the classical case. Corresponding to the above positive measure is the normalizing factor, customarily called the {\bf partition function}
\be
\mathcal Z_N:=  \int \int \d M_1 \d M_2 \frac {\alpha(M_1) \beta(M_2)}{\det(M_1+M_2)^N}
\ee 
whose dependence on the measures $\a,\beta$ carries all relevant information about the model.

We will show that this model is related to Cauchy biorthogonal polynomials
\be
\int_{\R_+}\int_{\R_+} p_n(x) q_m(y) \frac{\alpha(x)\d x\ \beta(y) \d y}{x+y} = \delta_{mn}
\ee
defined and studied in \cite{Paper1} in relation with the spectral theory of the cubic string and the Degasperis-Procesi wave equation (see also \cite{bss-moment, bss-stieltjes, ls-cubicstring, ls-invprob}).
Contrary to the biorthogonal polynomials of \cite{EynardMehta, BEH_dualityCMP, BEH_diffCMP} the properties of these polynomials do not depend on the measures $\alpha, \beta$ and the main highlights are 
\begin{enumerate}
\item they solve a four-term recurrence relation;
\item their zeroes are positive and simple;
\item their zeroes have the interlacing property;
\item they possess Christoffel--Darboux identities relevant to matrix models;
\item they can be characterized by a (pair of) Riemann--Hilbert problem(s) of size $3\times 3$;
\item the steepest descent method is fully applicable.
\end{enumerate}
Our recent paper  \cite{Paper1} contains a detailed discussion of all these points except the last one which will be addressed in a  forthcoming publication.

The paper is organized as follows; in Sect. \ref{CBOPs} we review our previous results \cite{Paper1} on the Cauchy biorthogonal polynomials and the relevant formul\ae\ and features needed in the following. In particular the Christoffell--Darboux identities (Sec.\ref{CDIs}), the  Riemann--Hilbert characterization in terms of $3\times 3$ piecewise analytic matrices (Sect. \ref{RHPS}). In Prop. \ref{Hkernel}  we introduce the $3\times 3$ {\em matrix kernel} in terms of the solution of the Riemann--Hilbert problems; this  will be used in later section to describe the spectral statistics of the two matrices.
In Sect. \ref{sect3} we introduce in detail the two--matrix model that we have outlined above and show how to reduce the study of its spectral statistics to Cauchy biorthogonal polynomials using a formula  appeared in \cite{HarnadOrlov}. We also indicate (without details) how to deal with a similar model where $R$ matrices are linked in a chain (Sect. \ref{Chain}).

In Sect. \ref{Diagram} we show how a formal treatment of the partition function of the model for large sizes of the matrices can be used to extract combinatorial information for certain bi-colored ribbon graphs (as it has been done for the Hermitean one-matrix model in  \cite{BIPZ, KazKosMig}  and for the multi-matrix models with exponential coupling in \cite{EynardLoop0}). 

In Sect. \ref{seclargen} we show that a saddle--point treatment of the partition function  (see for example \cite{diFrancesco} for the one-matrix case) leads to a  three--sheeted covering of the spectral plane (a {\em trigonal curve}); this  (pseudo) algebraic curve plays the same r\^ole as the hyperelliptic curve in the Hermitean matrix model. 

The Appendices are devoted to the relation between our proposed model and other matrix models with rectangular matrices (and possibly with Grassmann entries, App. \ref{rectangular}) and a connection (App. \ref{Onn} with the $O(1)$ model of self-avoiding loops (\cite{EynardOn} and references therein).
\br
For a particular case of measures $ \alpha(x) = x^a {\rm e}^{-x}$ and $\beta(y) = y^b {\rm e}^{-y}$ the corresponding biorthogonal polynomials appeared (in a somewhat disguised form) in the work \cite{Borodin1}. As observed therein they are related to the classical Jacobi orthogonal polynomials for the weight $x^{a+b}\d x$ on $[0,1]$. We thank A. Borodin for pointing out this connection.
\er
\section{Cauchy biorthogonal polynomials}
\label{CBOPs}
Let $K(x,y) = \frac 1{x+y}$ be the {\bf Cauchy kernel} on $\R_+\times \R_+$. It is known that it is totally positive in the sense of the classical definition \cite{gantmacher-krein,Karlin}: 

\bd
A {\bf totally positive kernel} $K(x,y)$ on $\mathfrak I\times \mathfrak J \subset \R\times \R$ is a function such that for all $n\in \N$ and ordered $n$-tuples $x_1<x_2<\dots<x_n$, $y_1<y_2<\dots <y_n$ we have the strict inequality
\be
\det[K(x_i, y_j)]>0\ .
\ee
\ed
It was shown in \cite{Paper1} that for any totally positive kernel $K(x,y)$ (hence also for the Cauchy kernel) and for any pair of measures $\alpha(x)\d x , \beta(y)\d y$ supported in $\mathfrak I, \mathfrak J\subset \R_+ $ respectively, the matrix of {\bf bimoments}
\be
I_{ij}  = \int_{\mathfrak I}\int_{\mathfrak J} x^i y^j K(x,y) \alpha(x)\d x \beta(y) \d y 
\ee
is a {\bf totally positive matrix}, that is, every square submatrix  has a positive determinant. 
This guarantees the existence of {\bf biorthogonal polynomials} $\{p_j(x),q_j(y)\}$ of exact degree $j$ such that 
\be
\int_{\mathfrak I}\int_{\mathfrak J} p_j( x)q_k( y) K(x,y) \alpha(x) \d x \beta(y) \d y = \delta_{jk}.
\ee
The polynomials  $p_j(x),q_j(y)$ are defined uniquely up to a $\C^\times$ action $p_j\mapsto \lambda_j p_j(x), q_j\mapsto \frac 1{\lambda_j} q_j$  and the ambiguity can be disposed of by requiring that their  leading coefficients are the same and positive. 
With this understanding we have $p_n(x) = \frac 1{c_n} x^n + \dots$ and $q_n = \frac 1{c_n} y^n + \dots$, with the positive constant $c_n$ given by  \cite{Paper1} 
\bea
c_n  = \sqrt{\frac {D_{n+1}}{D_n}}\ ,\qquad D_n:= \det\big[I_{jk}\big]_{0\leq j,k\leq n-1}\ ,\ \ 
I_{jk}:= \int_{\R_+}\int_{\R_+} x^j y^k K(x,y)\a(x)\d x \beta(y)\d y\ .\label{cn}
\eea

The determinantal expressions for the BOPs in terms of bimoments can be obtained using Cramer's rule \cite{Paper1}.
\bp[Thm. 4.5 in \cite{Paper1}]
For any totally positive kernel $K(x,y)$ the zeroes of the biorthogonal polynomials $p_j$'s ($q_j$'s) are simple, real and contained in the convex hull of the support of the measure $\alpha$ ($\beta$ respectively).
\ep

In  the case $K(x,y) = \frac 1{x+y}$ we named the corresponding polynomials $\{p_j(x),q_j(y)\}$  {\bf Cauchy BOPs} and proved, in addition, 
\bp[Thm. 5.2 in  \cite{Paper1}]
 The roots of adjacent polynomials in the sequences $\{p_j(x)\}$,  $\{q_j(y)\}$ are interlaced.
\ep

Cauchy BOPs enjoy more structure than the BOPs associated to a generic kernel: they solve a {\bf four term recurrence relation} of the form 
\bea
&\& x(\pi_{n-1} p_n(x) - \pi_{n} p_{n-1}(x)) = a_{n}^{(-1)} p_{n+1}(x) + a_{n}^{(0)}p_n(x) + a_{n}^{(1)} p_{n-1}(x) + a_{n}^{(2)} p_{n-2}(x)\nonumber\\
&\& y(\eta_{n-1} q_n (y)- \eta_{n} q_{n-1}(y))  = b_{n}^{(-1)} q_{n+1}(y) + b_{n}^{(0)} q_n(y) + b_{n}^{(1)} q_{n-1}(y) + b_{n}^{(2)} q_{n-2}(y)\label{recrels}
\eea
where
\be
\pi_n := \int_{\R_+} p_n(x)\a(x)\d x\ ,\qquad
\eta_n := \int_{\R_+}q_n(y)\beta(y)\d y\ .\label{pieta}
\ee
As proved in \cite{Paper1} $\pi_n, \eta_n$  are {\em strictly positive}.
The other coefficients appearing in the recurrence relations are described in loc. cit.

We also need to introduce certain auxiliary polynomials, $\wh q_n, \wh p_n$ whose defining properties are
 \begin{enumerate}
 \item $\deg \wh q_n = n+1$, $\deg \wh p_n = n$;
 \item $\ds \int \wh q_n \d \beta = 0$ or $\wh q_n (y)= \frac {q_{n+1}(y)}{\eta_{n+1}} - \frac {q_n(y)}{\eta_{n}}$;
\item $\ds \iint \wh p_n(x) \wh q_m(y) \frac {\d\alpha \d\beta}{x+y} = \delta_{mn}$ or $\frac 1{\eta_{n}} (\wh p_{n-1}-\wh p_{n}) = p_n$. 
\end{enumerate}
 In addition $\wh p_n, \wh q_n$ admit the determinantal representations:
 \bea
\wh q_n(y) &\&= \frac{1}{\eta_{n}\eta_{n+1}\sqrt{D_nD_{n+2}}} \det \le[
 \begin{array}{cccc}
 I_{00} & \dots &&I_{0n+1}\\
 \vdots& && \vdots\\
 I_{n-1\,0}& \dots&&I_{n-1\,n+1}\\
 \beta_0& \dots & &\beta_{n+1}\\
 1 & \dots &&y^{n+1} 
 \end{array}
 \ri]\\
\wh p_n(x) &\&   =\frac {1}{D_{n+1}} \det \le[
 \begin{array}{cccc}
 I_{00} & \dots &I_{0\,n}&1\\ 
 \vdots& && \vdots\\
 I_{n-1\,0}& \dots&I_{n-1\,n}&x^{n-1}\\
 I_{n 0}& \dots & I_{n\,n}&x^n\\
 \beta_0& \dots &\beta_{n}& 0 
 \end{array}
 \ri]\label{detwhp}
 \eea

After lengthy manipulations one obtains several Christoffel--Darboux--like identities (CDIs) 
for Cauchy BOPs which play a crucial r\^ole in what follows and hence will be carefully described. 

\subsection{Christoffel--Darboux Identities}
\label{CDIs}
Using  recurrence coefficients  featured in eqs. (\ref{recrels})
we define the following $3\times 3$ matrices
\be
\mathbb A_n(x) = \le[
\begin{array}{cc|c}
0&0&b_{n+1}^{(2)}\\
\hline
-b_{n-1}^{(-1)} & - b_{n}^{(0)}  + \frac x {\eta_{n+1}} & 0\\
0&-b_{n}^{(-1)}&0
\end{array}\ri]
\ ,\ \ 
\mathbb B_n(y) = \le[
\begin{array}{cc|c}
0&0&a_{n+1}^{(2)}\\
\hline
-a_{n-1}^{(-1)} & - a_{n}^{(0)}  + \frac  y{\pi_{n+1}} & 0\\
0&-a_{n}^{(-1)}&0
\end{array}\ri]
\label{CDIkernels}
\ee

In addition we define the following integral transforms
\bea
\label{eq:qs}
&& {q_{n}^{(1)}}(w) := \int_{\R_+} \frac {q_n(\zeta)\beta \d\zeta}{w-\zeta} \ ;\qquad 
 {q_n^{(2)}}(w):=  \int_{\R_+} \frac {q_n^{(1)}(-x)}{w+x}\a(x)\d x\\
&&   p_{n}^{(1)} (z) :=  \int_{\R_+} \frac {p_n(\xi)\alpha\d\xi}{z-\xi}\ ;\qquad
 {p_n^{(2)}}(z):=  \int_{\R_+} \frac {p_n^{(1)}(-y)}{z+y}\beta(y)\d y\ , \label{eq:ps}
 \eea
 and the following vectors 
 \be
 \vec \q_n^{(\mu)}(x)  = \le[
 \begin{array}{c}
 q_{n-2}^{(\mu)}(x)\\
 q_{n-1}^{(\mu)}(x)\\
 q_{n}^{(\mu)}(x)
 \end{array}\ri]\ ,\qquad 
  \vec \p_n^{(\mu)}(x)  = \le[
 \begin{array}{c}
 p_{n-2}^{(\mu)}(x)\\
 p_{n-1}^{(\mu)}(x)\\
 p_{n}^{(\mu)}(x)
 \end{array}\ri]
 \ee
 where $\mu=0,1,2$ and $q_n^{(0)} \equiv q_n$, $p_n^{(0)}\equiv p_n$.
 For $\alpha(x), \beta(y)$ we define $\alpha^\star(x) = \alpha(-x)$ and $\beta^\star(y) = \beta(-y)$; next we define the Weyl functions (or Markov-functions)  as 
\begin{align}
&W_{\beta}(z)=\int \frac{1}{z-y}  \beta(y)\d y = -W_{\beta^\star}(-z),  &W_{\alpha^*}(z)&=\int \frac{1}{z+x} \alpha(x)\d x = -W_\a(-z), \cr
&W_{\alpha^*\beta}(z)=-\iint \frac{1}{(z+x)(x+y)}\alpha(x) \beta(y)\d x \d y
&W_{\beta \alpha^*}(z)&=\iint \frac{1}{(z-y)(y+x)} \alpha(x) \beta(y)\d x \d y   
\end{align}

\bp[Thm 7.3 in \cite{Paper1}]
\label{propCDI}
The following Christoffel--Darboux-like identities hold
\bea
&\&(z+w)\sum_{j=0}^{n-1} q_j^{(\mu)}(w) p_j^{(\nu)}(z)  ={\vec \q_n^{(\mu)} (w) \cdot \mathbb A(-w)\cdot  \wh{\p_n}^{(\nu)}(z)} - \mathbb F(w,z)_{\mu,\nu}
\label{CDI_mu_nu}\\
&\& \mathbb F(w,z) =  \begin{bmatrix}
0&0&1\\
0& 1&W_{\beta^*}(z)+W_{\beta}(w) \\
1&W_{\alpha}(z)+W_{\alpha^*}(w)&W_{\alpha^*}(w)W_{\beta^*}(z)+W_{\alpha^*\beta}(w)+W_{\beta^*\alpha}(z)  
\end{bmatrix}
\eea
where the auxiliary vectors marked with a hat are characterized by a Riemann--Hilbert problem described below.
\ep
\bc [Thm 7.4 in \cite{Paper1}]
\label{cordual}
Evaluating (\ref{CDI_mu_nu}) on the ``antidiagonal'' $z=-w$ gives the {\bf perfect duality} 
\bea
{\vec \q_n^{(\mu)} (w) \cdot \mathbb A(-w)\cdot  \wh{\p_n}^{(\nu)}(-w)} = \mathbb J_{\mu\nu}\cr\cr
\mathbb J:= \le[
\begin{array}{ccc}
0&0&1\\ 0&1&0\\
1&0&0
\end{array}
\ri]
\eea
\ec
\br
The proposition above defines, in fact,  $9$ identities, but we will only need the $4$ identies corresponding to $\mu,\nu=0,1$ (i.e. the principal submatrix of size $2\times 2$).
\er
\subsection{Riemann--Hilbert characterization of the integrable kernels}
\label{RHPS}
The sums appearing on the left hand side in Prop. \ref{propCDI} are all examples a general framework of ``integrable kernels'' that were studied in great generality in \cite{IIKS1, IIKS2, IIKS3, IIKS4, HarnadIts}. 

%
\bp[ Prop. 8.1 in \cite{Paper1}]
\label{RHP1}
Consider the Riemann--Hilbert problem (RHP) of finding a matrix $\Gamma(w)$ such that 
\begin{enumerate}
\item $\Gamma(w)$ is analytic on $\C \setminus (supp(\beta) \cup supp (\alpha^\star))$
\item $\Gamma(w)$ satisfies the jump conditions 
\be \label{eq:RHq}
\begin{split}
\Gamma(w)_+ & = \Gamma(w)_-\le[\begin{array}{ccc}
1 &  -2\pi i \beta & 0 \cr
0&1&0\cr
0&0&1
\end{array}\ri]\ , \qquad w\in supp(\beta)\subset \R_+\cr
\Gamma(w)_+ & = \Gamma(w)_- \le[
\begin{array}{ccc}
1&0&0\\
0&1&  -2\pi i  \alpha^*\\
0&0&1
\end{array}
\ri]\ ,\qquad w\in supp( \alpha^*)\subset \R_-  
\end{split}
\ee
\item its asymptotic behavior at $w=\infty$ $\Im(w)\neq 0$ is 
\bea \label{eq:Gamma-as}
\Gamma(w)  = (\1 + \mathcal O(w^{-1}))\le[
\begin{array}{ccc}
w^n& 0 & \\
0& w^{-1}& 0 \\
0&0&w^{-n+1}
\end{array}
\ri]
\eea
\end{enumerate}
Then such a $\Gamma(w)$ exists and is unique. Moreover $\Gamma(w)$ can equivalently be written as:
\begin{align}\label{eq:q-recovery}
\Gamma(w)=\begin{bmatrix}
c_n\eta_n&0&0\\
0&\frac{1}{\eta_{n-1}}&0\\
0&0&\frac{(-1)^{n-1}\eta_{n-2}}{c_{n-2}}
\end{bmatrix}
\begin{bmatrix}
\wh q_{n-1}^{(0)}& \wh q_{n-1}^{(1)}&\wh q_{n-1}^{(2)}\\
q_{n-1}^{(0)}&q_{n-1}^{(1)}&q_{n-1}^{(2)}\\
\wh q_{n-2}^{(0)}& \wh q_{n-2}^{(1)}&\wh q_{n-2}^{(2)}
\end{bmatrix}. \end{align}
or also 
\be
\Gamma(w) :=\begin{bmatrix}
1&-c_n \eta_n&0\\
0&1&0\\
0&(-1)^{n-1}\frac{\eta_{n-2}}{c_{n-2}}&1 \end{bmatrix}
\le[\begin{array}{ccc}
0 &0&c_n
\cr 
0 &\frac{1}{\eta_{n-1}}&0
\cr
\frac {(-1)^{n}} {c_{n-2}}&0&0
\end{array}\ri] 
\overbrace{ [\vec \q_{_n}^{(0)}(w), \vec \q_{_n}^{(1)}(w), \vec \q_{_n}^{(2)}(w)]}^{\ds :=Y(w)}
\label{normalizedqRHP}
\ee
where the normalization constants $\eta_n, c_n$ have been introduced in (\ref{cn}, \ref{pieta}).
\ep
\br
The quantities referring to the letter $q$ on the right hand side in Prop. \ref{propCDI} can be extracted from the RHP involving $\Gamma$. The next proposition will achieve the same goal for the remaining quantities.
\er

\bp[Prop. 8.2 in \cite{Paper1}]
\label{RHP2}

Consider the Riemann--Hilbert problem (RHP) of finding a matrix $\wh \Gamma(z)$ such that 
\begin{enumerate}
\item $\wh \Gamma(z)$ is analytic on $\C \setminus (supp(\alpha) \cup supp (\beta^\star))$
\item $\wh \Gamma(z)$ satisfies the jump conditions 
\be \label{eq:RHphat}
\begin{split}
\wh \Gamma(z) _+ &= \wh \Gamma(z)_- \le[
\begin{array}{ccc}
1 &   -2\pi i \alpha(z)  & 0\cr
0&1&0\\
0&0&1
\end{array}
\ri]\ ,\ \ z\in supp(\alpha)\subseteq \R_+\\
\wh \Gamma(z) _+ &= \wh \Gamma(z)_- \le[
\begin{array}{ccc}
1 &  0 & 0\cr
0&1& -2\pi i  \beta^*\\
0&0&1
\end{array}
\ri]\ ,\ \ z\in supp( \beta^*)\subseteq \R_-, 
\end{split}
\ee 
\item its asymptotic behavior at $z=\infty$ $\Im(z)\neq 0$ is 
\bea \label{eq:Gammahat-as}
\wh \Gamma(z) =  \le(\1  + \mathcal O\le(\frac 1 z\ri)\ri)\le[
\begin{array}{ccc}
z^n & 0& \cr
0 &1 &0\cr
0 &0& \frac 1{ z^{n}} 
\end{array} 
\ri].  \label{RHhatasymp}
\eea

\end{enumerate}
Then such a $\wh \Gamma(z)$ exists and is unique. Moreover $\wh \Gamma(z)$ can equivalently be written as:
\begin{align}
\label{eq:p-recovery}
\wh \Gamma(z)=\begin{bmatrix}c_n&0&0\\0&-1&0\\0&0&\frac{(-1)^n}{c_{n-1}}\end{bmatrix}
\begin{bmatrix}p_{0,n}&p_{1,n}&p_{2,n}\\
\wh p_{0,n-1}&\wh p_{1,n-1}&\wh p_{2,n-1}\\p_{0,n-1}& p_{1,n-1}&p_{2,n-1}\end{bmatrix}.
 \end{align}
or also 
\be
\wh \Gamma(z) =
 \le[
\begin{array}{ccc}
0&0& -\frac {c_n}{\eta_n} \\
0&-1&0\\
\frac {(-1)^{n}}{c_{n-1}\eta_{n-1}} &0&0
\end{array}
\ri]\le[
\begin{array}{crc}
1 & -1 & 0\\
0 &1&0\\
0&-1 &1
\end{array}
\ri] 
 \overbrace{\le[\vec{\wh \p}_{_n}^{(0)} (z),\vec{\wh \p}_{_n}^{(1)} (z),\vec{\wh \p}_{_n}^{(2)} (z)\ri]}^{\ds := \wh Y(z)}.
\label{normalizedphatRHP}
\ee 
where the normalization constants $\eta_n, c_n$ have been introduced in (\ref{cn}, \ref{pieta}).
\ep
\br
These polynomials $\wh p_n(z):= \wh p^{(0)}_n(z)$ and the auxiliary functions $\wh p_n^{(1)}(z), \wh p_n^{(2)}(z)$  were introduced in \cite{Paper1} independently from a RHP formulation, but for all practical purposes the formulation above is sufficiently explicit.
\er


Finally the RHPs above allow us to reconstruct the ratio of two consecutive principal minors of the bimoment matrix (this will become relevant when discussing the partition function of the matrix model)
\bp[Corollary 8.1 in \cite{Paper1}]
\label{corratio} If  $D_n$ is a leading principal $n\times n$ minor of the bimoment matrix $I$, then
\be
\frac {D_{n}}{D_{n-1}} = (-1)^n \lim_{w\to\infty} w^{2n-1} \frac{\Gamma_{2,3}(w)}{\Gamma_{2,1}(w)}= c_{n-1}^2 >0
\ee
\ep

The CDIs  in Prop. \ref{propCDI} can be written in a very simple form in terms of the solutions of the Riemann--Hilbert problems. To show this let us momentarily denote by $Y(w)$ and $\wh Y(z)$ the matrices  with columns given by the $\vec \q_n^{(\mu)}(w)$ and $\vec \p_n^{(\nu)}(z)$ respectively. Then (\ref{propCDI}) can be reformulated as 
\be
(z+w)  \mathbb H_n(z,w) := (z+w) \le[\sum_{j=0}^{n-1} q_j^{(\mu)}(w) p_j^{(\nu)}(z)\ri]_{\mu,\nu}  +  \mathbb F(w,z)= Y^t(w) \mathbb A(-w)\cdot \wh Y(z)
\ee
Now the perfect duality of Cor.  \ref{cordual} implies that 
\be
Y(w)\mathbb A(-w) = \mathbb J \wh Y(-w)^{-1}\ .
\ee
Note that the solutions of the RHPs $\Gamma(w), \wh \Gamma(z)$ differ from  $Y(w), \wh Y(z)$ only by some constant invertible {\em left} multipliers, and hence 
\bea
Y(w)\cdot \mathbb A(-w)\cdot\wh Y(z) =  \mathbb J\cdot  \wh Y(-w)^{-1}\cdot\wh Y(z) = \mathbb J \cdot \wh \Gamma(-w)^{-1} \cdot \wh \Gamma(z)\ .
\eea
We collect this into the following proposition for later reference
\bp
\label{Hkernel}
The {\bf matrix kernel} $\mathbb H_n(z,w)$
\be
[\mathbb H_{n}]_{\mu \nu} (z,w)  :=\le( \sum_{j=0}^{n-1} q_j^{(\mu)}(w) p_j^{(\nu)}(z) +\frac  {  \mathbb F_{\mu\nu}(w,z)} {z+w} \ri)
\ee
is given in terms of the solution of the 
Riemann--Hilbert  problem (\ref{normalizedphatRHP}) - (\ref{RHhatasymp}) as 
\be
\mathbb H_n(z,w) := \frac{ \mathbb J\cdot \wh \Gamma(-w)^{-1}\cdot \wh \Gamma(z) }{z+w}
\ee
\ep
The relevance of the matrix kernel $\mathbb H_n(z,w)$ will become clear when we will discuss the spectral statistics of the matrix model.

\section{Matrix Models}
\label{sect3}
Consider the vector space $\mathcal H_N$ of Hermitean matrices of size $N$ endowed with the $U(N)$--invariant Lebesgue measure
\be
\d M := \prod_{i<j} \d \Re(M_{ij}) \d \Im (M_{ij}) \prod_j \d M_{ii}\ .
\ee
Inside $\mathcal H_N$ we consider the {\bf convex cone} of {\bf positive definite}  matrices
\be
\mathcal H_{N}^+ := \{M = M^\dagger, M>0\}
\ee
with the induced measure. Let $\alpha, \beta$ be two positive densities of finite mass supported on the positive real axis: 
$\alpha(M)$ will simply mean the product measure on the {\bf eigenvalues} of $M$. 


Define now the measure on $\mathcal H_N^+ \times \mathcal H_N^+ = \{ (M_1,M_2) \}$
\be
\d\mu(M_1,M_2) := \d M_1 \d M_2 \frac {\alpha(M_1) \beta(M_2)}{\det(M_1+M_2)^N}  
\label{mumeas}
\ee

\bd
For the positive finite--mass measure (\ref{mumeas}) we define the {\bf partition function} as the integral 
\be
\mathbf  Z_{N} := \int_{\mathcal H_N^+\times \mathcal H_N^+} \d\mu(M_1,M_2)
\ee
\ed 

The resulting random matrix model falls into the general class of {\em two--matrix--models} although the ``coupling'' term is not the most common one (which is ${\rm e}^{Tr(M_1M_2)}$). However these and much more general  couplings have been considered relatively recently in \cite{HarnadOrlov}.

As customary,  we re-express the Lebesgue measures $\d M_i$ in terms of the {\em normalized}  Haar measure $\d U_i$ of the unitary group $U(N)$ and the Lebesgue measure on the cone $\R_+^N$
\be
\d M_1  =G_N \d U \prod \d x_j \Delta^2(X)\ ,\qquad
\d M_2 = G_N\d V \prod \d y_j \Delta^2(Y)\ ,
\ee
where $\Delta(X):= \det [x_i^{j-1}]$ is the Vandermonde determinant
associated with the $n$-tuple $X$  of ordered eigenvalues $x_1\leq \ldots\leq x_n$ of $M_1$ and 
$\Delta(Y)$ is defined in the same way relative to the matrix $M_2$. The constant $G_N$ is not of much relevance. It  depends only on $N$ but not on the densities.
The resulting measure involves two copies of the unitary group and one of the integrals can be performed leading (up to a multiplicative constant depending on the normalizations of the Haar measures and the size $N$) to the measure below on the spectra of $M_1, M_2$
\bea
\d\mu(X,Y)  =G_N^{2}\Delta^2(X) \Delta^2(Y) \le(\int_{U(N)}\frac {\d U}{\det(X + UYU^\dagger)^N} \ri)
\alpha(X)\d X 
\beta(Y) \d Y \\
\alpha(X) := \prod_{j=1}^{N} \alpha(x_j)\ ,\  \d X :=\prod_{j=1}^N\d x_j \qquad
\beta(Y) := \prod_{j=1}^N \beta(y_j)\ ,\ \d Y:=\prod_{j=1}^N\d y_j \ .
\eea
At this point we need to compute the integral over $U(N)$;  we can use the result on pages 23-24 of \cite{HarnadOrlov} for the special example (A-28)
\bea
\int_{U(N)} \det(\1 - z\, A U B U^\dagger) ^{-r}\d U  = C_{N,r} \frac {\det \le[ (1-z a_i b_j)^{N-r-1}\ri]_{1\leq i,j\leq N}} {\Delta(A) \Delta(B)}
\label{38}
\\
C_{N,r} := \frac {\prod_{k=1}^{N-1} k!} {z^{N(N-1)/2} \prod_{k=1}^{N-1} (r-N+1)_k}
\eea
where 
\be
(r-N+1)_k := \prod_{j=1}^{k} (r-N+j)
\ee
is the Pochammer symbol.

Setting $z=-1,\  A = X^{-1},\ B = Y$ in (\ref{38}) we obtain
\be
\int_{U(N)} \frac {\d U}{\det (\1 + X^{-1}  U Y U^\dagger) ^r}  = C_{N,r} \frac {\det \le[ (1+ \frac  {y_j}{x_i})^{N-r-1}\ri]_{1\leq i,j\leq N}} {\Delta(X^{-1}) \Delta(Y)}
\ee
Multiplying both sides by $\det(X)^{-r}$ we then obtain
\be
\int_{U(N)} \frac {\d U}{\det(X + U Y U^\dagger) ^r}  =(-1)^{N(N-1)/2}  C_{N,r} \frac {\det \le [ (x_i+y_j)^{N-r-1}\ri]_{1\leq i,j\leq N}} {\Delta(X) \Delta(Y)}
\ee

The case of main relevance to us is $r=N$, 
which yields ($C_{N,N} = (-1)^{N(N-1)/2}$) 
\be
\int_{U(N)} \frac {\d U}{\det(X +   U Y U^\dagger) ^N}  =  \frac {\det \le[ \frac 1{x_i+ y_j}\ri]} {\Delta(X) \Delta(Y)}
\ee
This shows that the measure $\d\mu(M_1,M_2)$ can be reduced to a measure on the spectra of the two matrices (we use the same symbol for the measure on the eigenvalues)
\bea
\d\mu(X,Y) = G_{N}^2 \Delta^2(X) \Delta^2(Y)\frac { \det[K (x_i,y_j) ]}{\Delta(X) \Delta(Y)} \alpha(X) \d X \beta(Y)\d Y \cr
K(x,y) = \frac 1{x+y}\ .
\eea

In fact we could have used the general formula (\ref{38}) for any $r := N+h$;
note, however, that for $h$ integer and less than $-1$ Harnad--Orlov's formula in the form presented above cannot be used, since the determinant in the numerator vanishes (for $N\geq -h$) and so do some denominators in the definition of the constants $C_{N,r}$. In other words, one should take an appropriate limit and use de l'H\^opital's rule.
For any $h\geq 0$ or any $h\notin -\N$, tracing the steps above 
 one could obtain general models where the reduced measures on the spectra have a form 
\bea
\d\mu_h	(X,Y) = G_N^2 C_{N,r}\Delta^2(X) \Delta^2(Y)\frac { \det(K_h (x_i,y_j) )}{\Delta(X) \Delta(Y)} \alpha(X) \d X \beta(Y)\d Y \cr
K_h(x,y) = \frac 1{(x+y)^{1+h}}\ ,
\eea
corresponding to the unreduced measure 
\be
\d\mu_h (M_1,M_2) = \d M_1\d M_2 \frac {\alpha(M_1)\beta(M_2)}{\det(M_1 + M_2)^{N+h}}\ .\label{muh}
\ee

It will be important for us in what follows that for any value of $h\not\in -\N$  the kernel $K_h$ is totally positive or at least {\em sign regular} on
$\mathbb{R}_+\times \mathbb{R}_+$.  (Sign-regularity means that determinants in Def. 2.1 are nonzero and their sign depends only on their size.)

\bl
Consider  the kernel $K_h=\frac 1{(x+y)^{1+h}}$ ,
 restricted to $(x,y)\in \mathbb{R}_+\times \mathbb{R}_+$. 
 Then, 
 \begin{enumerate}
 \item for $1+h >0, \, K_h$  is totally positive,
 \item for $1+h \in \R_-\setminus -\N, \, K_h$ is sign-regular.  
 \end{enumerate}
 \el
 {\bf Proof.}
To prove this assertion  define for $0 <  y_0 < \cdots  < y_n$ functions
$k_i(x) = \frac 1{(x+y_i)^{s}}\ , \ i=0, \ldots,  n, \ s = 1 + h$ . Denote for $j=1, 2, ...$, $c_j(s)= s (s+1)\ldots (s+j-1)$ and $c_0(s)=1$. Then the Wronskian $W(k_0, \ldots, k_n)(x)$ 
of $k_0, \ldots, k_n$ is equal to
\be
\det \left  [  (-1)^{j} c_j(s) \frac 1{(x+y_i)^{s+j}} \right  ]_{i,j=0}^n = (-1)^{\frac{n(n-1)}{2}}\prod_{j=0}^n \frac {c_j(s)} {(x+y_i)^s}  \det\left  [   \frac 1{(x+y_i)^{j}} \right  ]_{i,j=0}^n
\ee
Thus, $W(k_0, \ldots, k_n)(x)$  is a nonzero multiple of $ (-1)^{\frac{n(n-1)}{2}} \Delta(z_0, \ldots, z_n)$, where $z_i= \frac 1{(x+y_i)} $ and $\Delta(z_0, \ldots, z_n)$ is the Vandermonde determinant constructed out of the $z_i$'s. Since $z_0 > \ldots > z_n$, we see that 
\be
 (-1)^{\frac{n(n-1)}{2}} \Delta(z_0, \ldots, z_n) =  \Delta(z_n, \ldots, z_0) > 0
\ee
 Thus $W(k_0, \ldots, k_n)(x)$ is {\bf positive} for any $n$ if  $s > 0$ since all the $c_j(s)$ are positive numbers; if $s<0$ then -  denoting by $[s]$ the greatest integer less than $s$ (hence negative) - we see that the sign of $W$ is 
\be
sign(W) = (-1)^{-[s]n - \frac {[s](1-[s])}2}
\ee
 for any $n\geq 0$ and $x \geq 0$. Together these observations imply
 (Thm 2.3, ch. 2 of \cite{Karlin}), that $K_h$ is totally positive for $s> 0$ on
$\mathbb{R}_+\times \mathbb{R}_+$ or at least sign regular for $s\in \R_-\setminus -\N$.
{\bf Q.E.D.}\par \vskip 5pt

%

In App. \ref{rectangular} we will show that for $h$ integer or half--integer, the model (\ref{muh}) is the reduction of a $3$--matrix model, where a ``ghost'' gaussian matrix $A$ has been integrated out. Depending on the range $h<N$ or $h>N$ the matrix $A$ consists of ordinary variables (bosons) or Grassmann variables.

\br
The fact that the measure $\d \mu_h$ depends on $N$ (in the determinant in the denominator) is a  feature of the model rather than a problem; indeed, in studying the large $N$ limit it is natural to make the strength of the interaction increase at the same rate as  the size of the matrices. For example, in the ``standard'' two--matrix model one considers the interaction ${\rm e}^{c N Tr( M_1M_2)}$
\er
\subsection{Correlation functions: Christoffel--Darboux kernels}
In this section we will compute the correlation functions of the model. More precisely,

\bd The correlation functions of the model are defined as
 \bea
&\& \mathcal R^{(r,k)} (x_1,\dots x_r; y_1,\dots, y_k) :=
\cr &\& \frac {N!\prod_{j=1}^r \alpha(x_j) \prod _{j=1}^k \beta(y_j)}{(N-r)!(N-k)!\mathcal Z_N}
\int \prod _{\ell = r+1}^N \alpha(x_j)  \d x_\ell \prod_{j=k+1}^N\beta(y_j) \d y_j  \Delta^2(X) \Delta^2(Y)\frac { \det[K (x_i,y_j) ]}{\Delta(X) \Delta(Y)} 
\eea
where $\mathcal Z_N:=\frac 1 {N!} \int \Delta(X) \Delta(Y) \det [K(x_i,y_j)] _{i,j\leq N}\alpha(X) \beta(Y)\d X\d Y$.
\ed
These functions allow one to compute the probability of having $r$ eigenvalues of the first matrix and $k$ eigenvalues of the second matrix in measurable sets of the real axis.

The  computations of these correlation functions in term of biorthogonal functions reported below follows  the general approach in  \cite{EynardMehta} and \cite{Harnad}.

Using the well--known formula for the Cauchy determinant 
\be
\det\le[\frac 1{x_i  + y_j}\ri] = \frac {\Delta(X) \Delta(Y)}{\prod_{i,j} (x_i + y_j)}
\ee
we obtain the  measure 
\be
\frac {\Delta(X)^2 \Delta(Y)^2}{\prod_{i,j} (x_i + y_j)} \alpha(X) \beta(Y)\d X \d Y\label{Cauchybello}
\ee
We will be using the correlation functions only to compute expectations of spectral functions, namely functions of $X, Y$ which are separately symmetric in the permutations of the $x_j$'s or $y_j$'s.
\begin{lemma} \label{lem:symintegral}
Suppose $F(X)$ is a symmetric function 
under  the action of the symmetric group $S_N$.
Then 
\begin{align*}
\frac 1 {N!}\int F(X) \Delta(X)  \det[K(x_j,y_j) ]\alpha(X)\d X
=
\int F(X) \Delta(X) \prod_{j=1}^{N} \frac {1 } {x_j + y_j}  \alpha(X)\d X 
\end{align*}
\end{lemma}
{\bf Proof.}
Under the assumption $F(X)= F(X_\sigma)$ for any $\sigma \in S_N$ we have 
\begin{align*} 
&\int F(X) \Delta(X)\prod_{j=1}^{N} \frac {1} {x_j + y_j} \alpha( X)\d X  = \\
& =\frac 1 {N!} \sum_{\sigma\in S_N} \int F(X) \Delta(X_\sigma) \prod_{j=1}^{N} \frac {1} {x_{\sigma(j)}+ y_j} \alpha( X)\d X=\cr
 & =\frac 1 {N!}\int F(X) \Delta(X)  \le( \sum_{\sigma\in S_N} \epsilon(\sigma) \prod_{j=1}^N K(x_{\sigma(j)},y_j) \ri)\alpha( X)\d X=\\&
 =\frac 1 {N!}\int F(X) \Delta(X)  \det[K(x_j,y_j) ]\alpha( X)\d X\ .  
\end{align*}
{\bf Q.E.D.}\par \vskip 5pt

\bc 
\label{cor:symm}
Let $F(X,Y)$ be symmetric with respect to either set of variables $X$ or $Y$.  
Then
\be 
 \frac 1 {N!}\int F(X,Y) \Delta(X)\Delta(Y)  \det[K(x_j,y_j) ]\alpha( X)\beta(Y) \d X\d Y = \int F(X,Y) \Delta(X)\Delta(Y) \prod_{j=1}^{N} \frac {1 } {x_j + y_j} \alpha(X) \beta(Y) \d X \d Y 
\ee 
\ec
Since we are interested in the {\em unordered} spectrum of the matrices $M_1,M_2$, in view of the above Cor. \ref{cor:symm} we will focus henceforth on the following unnormalized measure 
\be
\d \wt \nu(X,Y):= \Delta(X) \Delta(Y) \prod_{j=1}^{N} \frac {\alpha(x_j) \beta(y_j)} {x_j + y_j} \d X \d Y   = \Delta(X)\Delta(Y) \frac {\alpha(X)\beta(Y)}{\det(X+Y)} \d X \d Y 
\label{reducedmeasure}
\ee

From the properties of the Vandermonde determinant we can write 
\be
\Delta(X)\Delta(Y) =  \det\le[\wt p_i(x_j)\ri]\det\le[\wt q_i(y_j)\ri] 
\ee
where $\wt p_i, \wt q_i$ are any {\em monic} polynomials of exact degree $i$ in the respective variables ($i=0,\ldots,N-1$).

It is therefore natural \cite{EynardMehta} to choose the sets of monic
polynomials $\{\wt p_j(x), \wt q_j(y)\}_{j\in \N}$ to be  {\bf
  biorthogonal} with respect to the bi-measure $
\frac {\alpha(x) \beta(y)}{x+y} \d x \d y$.
This means that 
\be
\int_{\R_+}\int_{\R_+} \wt p_k(x)\wt q_\ell(y)\frac {\alpha(x) \beta(y)}{x+y} \d x \d y =  {c_k}^{2} \delta_{k\ell}.
\ee
The constant $c_k$ was defined  in (\ref{cn}, with $K(x,y)=1/(x+y)$).
The normalization constant for $\d \wt \nu$ is thus
\be
\mathcal Z_N =\iint \d\wt \nu(X,Y) =   N! \le(\prod_{k=0}^{N-1} c_{k}^2\ri)   = N! \det\big[I_{jk}\big]_{0\leq j,k\leq N-1}\label{partminor}
\ee

If we introduce the {\bf orthonormal biorthogonal polynomials}
\be
p_n:= \frac 1{c_n} \wt p_n\ ,\ \ \ \ q_n:= \frac 1{c_n} \wt q_n
\ee
then we can write the {\em normalized} measure as 
\be
\nu(X,Y) = \frac 1 {N!}  \det\le[ p_i(x_j)\ri]\det\le[ q_i(y_j)\ri]  \prod_{j=1}^{N} \frac {\alpha(x_j) \beta(y_j)} {x_j + y_j} \d^N X \d ^NY.\label{numeas2}
\ee

Notice now that the product of determinants in (\ref{numeas2}) is a determinant of the product of the two matrices indicated; hence 
\bea
\det\le[p_{i-1}(x_j)\ri]_{i,j\leq N}\det\le[q_{j-1}(y_i)\ri]_{i,j\leq N}  = \det \mathbb K_N(x_i, y_j)\\
 \mathbb K_N(x,y) := \sum_{j=0}^{N-1}p_j(x) q_j(y)
\eea
The kernel  $\mathbb K_N(x,y)$ is a ``reproducing'' kernel 
\be
\mathbb K_N(x,y) = \iint \mathbb K_{N} (x,z) \mathbb K_N(w,y) \frac { \alpha(w)\beta(z)\d w \d z}{z+w}
\ee
which follows immediately from the biorthogonality. In addition we have 
\be
\iint \mathbb K_{N} (x,y) \frac {\a(x)\beta(y)\d x \d y}{x+y} = N \ . 
\ee
Summarizing, in the new notation we have

\bp \label{lem:niceformulas}
The  probability measure on $X \times Y$  induced by (\ref{mumeas}) is given by: 
\be
\frac{1}{(N!)^2} \det[\mathbb K_N(x_i,y_j)] \det [K(x_i,y_j)]\alpha(X) \beta(Y)\d X \d Y 
\ee
while the correlation functions are: 
\begin{align}
&\mathcal R^{(r,s)}(x_1,\dots x_r; y_1,\dots, y_s) =\\
&\frac {\prod_{j=1}^r \alpha(x_j) \prod _{j=1}^s \beta(y_j) }{(N-r)!(N-s)!} \int  \det[\mathbb K_N(x_i,y_j)]\det[K (x_i,y_j)]\prod _{\ell = r+1}^N \alpha(x_j)\d x_j  \prod_{j=s+1}^N\beta(y_j)\d y_j 
\end{align}
\ep 

\bx 
\label{ex:1} Consider $r=1, s=0$.  In this case, with the help of Lemma \ref{lem:symintegral}, we obtain:
\be
\begin{split}
&\mathcal R^{(1,0)}=\frac{1}{(N-1)! N!}\alpha(x_1) 
\int \det[\mathbb K_N(x_i,y_j)]\det[K(x_i,y_j)] \prod_{\ell =2}^N\alpha(x_{\ell})\d x_\ell   \prod_{j=1}^N\beta(y_j)\d y_j=\\ 
&\frac{1}{(N-1)!} \alpha(x_1) \int \det[\mathbb K_N(x_i,y_j)]\prod_{i=1}^N \frac{1}{x_i+y_i}  \prod_{\ell =2}^N\alpha(x_{\ell}) \d x_\ell   \prod_{j=1}^N\beta(y_j)\d y_j=\\
&\frac{1}{(N-1)!} \alpha(x_1)
\sum_{\sigma, \sigma ' \in S_N} \epsilon(\sigma) \epsilon(\sigma')
\int p_{\sigma(1)-1}(x_1)\dots p_{\sigma(N)-1}(x_N)\\
& q_{\sigma'(1)-1}(y_1)\dots q_{\sigma'(N)-1}(y_N)\prod_{i=1}^N \frac{1}{x_i+y_i}  \prod_{\ell =2}^N\alpha(x_{\ell})\d x_\ell   \prod_{j=1}^N\beta(y_j)\d y_j= \alpha(x_1)\int \frac{\mathbb K_N(x_1,y_1)} {x_1+y_1}  \beta(y_1)\d y_1 
\end{split}
\ee  
\ex 
\subsubsection{Correlation functions in terms of biorthogonal polynomials}
In the paper by Eynard and Mehta \cite{EynardMehta} (generalized in \cite{Harnad})  they never used any specific information about the model they were considering (with the Itzykson--Zuber interaction) but only the fact that the matrices were coupled in a chain. 
We recall the relevant result here\footnote{It could be extended to the chain of matrices (see below) but it  becomes a bit cumbersome to describe.}. 
Define
\bea
&& H_{00}(x,y):= \mathbb K_N(x,y) \cr
&& H_{01}(x,x'):= \int H_ {00}(x,y)\frac {\beta(y) \d y}{x'+y}\ ,\qquad
 H_{10}(y,y'):= \int H_{00}(x,y') \frac{\alpha(x) \d x}{x+y}\cr
&& H_{11}(y,x):= \iint H_{00}(z,w)\frac{\alpha(z) \d z \beta(w) \d w}{(z+y)(x+w)} - \frac1{x+y}\ .\label{Hkernels}
\eea
Since  $H_{00}$ is a reproducing kernel
\be
\int H_{00}(x,y) H_{11}(y,x') \beta(y)\d y = H_{01}(x,x') - H_{01}(x,x') = 0
\ee
and similarly 
\be
\int H_{11}(x,z) H_{00}(z,x') \alpha(z) \d z = H_{10}(x,x') - H_{10}(x,x') \equiv 0\ .
\ee
Integrating these two equations against $(x+y')^{-1}\alpha(x) \d  x$ (the first) or $(x'+y')^{-1}  \beta(y')\d y'$ (the second) we find also 
\be
\int H_{11}(y,x)H_{10}(y,y')  \beta(y) \d y = \int 
\int H_{11}(x,y) H_{01}(y,y')  \alpha(y) \d y  \equiv 0
\ee

The correlation functions for $r$ eigenvalues $x_1,\dots, x_r$ of $M_1$ and $s$ eigenvalues $y_1,\dots, y_s$ of $M_2$ were computed in \cite{EynardMehta} and are given by 
\bea
\mathcal R^{(r,s)}(x_1,\dots, x_r; y_1,\dots, y_s) = \prod_{j=1}^r \alpha(x_j)\ \prod_{k=1}^s \beta(y_k) \times \cr
\times \det\le[
\begin{array}{c|c}
\big[H_{01}(x_i,x_j)\big]_{1\leq i,j\leq r} & \big[H_{00} (x_i, y_j) \big]_{1\leq i\leq r, 1\leq j\leq s}\\[10pt]
\hline
\rule{0pt}{16pt}\big[H_{11}(y_i,x_j)\big]_{1\leq i\leq s, j\leq r} & \big[H_{10} (y_i, y_j) \big]_{1\leq i,j \leq s}
\end{array}
\ri]
\eea
where $0\leq r \leq N$, $0\leq s \leq N$, $1\leq r+s$, with the understanding that if either $r$ or $s$ is $0$ then the corresponding blocks labeled by $r, \ s$ respectively, are absent. Thus, for example $\mathcal R^{(1,0)}(x_1) = \alpha(x_1) H_{0,1}(x_1,x_1)$. 

This is a nontrivial result and perhaps one can get a bit of an insight by 
considering the special case $r=s=N$.  
We know that in this case 
\begin{align}
&\mathcal R^{(N,N)}(x_1,\dots x_N; y_1,\dots, y_N) =\\
&\det[\mathbb K_N(x_i,y_j)]\det[K (x_i,y_j)]\alpha(X)  \beta(Y) 
\end{align}
so we have the following not at all obvious identity: 
\bl 
\be 
\det[K_N(x_i,y_j)]\det[K (x_i,y_j)]=\det\le[
\begin{array}{c|c}
\big[H_{01}(x_i,x_j)\big]_{i,j\leq N} & \big[H_{00} (x_i, y_j) \big]_{i, j\leq N}\\[10pt]
\hline
\rule{0pt}{16pt}\big[H_{11}(y_i,x_j)\big]_{i, j\leq N} & \big[H_{10} (y_i, y_j) \big]_{i,j \leq N}
\end{array}
\ri]
\ee
\el 
{\bf Proof}
It suffices to observe that the $2N \times 2N$ matrix on the right hand side has the following block structure: 
\be \label{def:matrixK}
\mathcal{K}=\begin{bmatrix} -\mathbf {P_0(x)}^T \mathbf {Q_1(-x)}& \mathbf {P_0(x)}^T \mathbf {Q_0(y)}\\
\mathbf {P_1(-y)}^T  \mathbf {Q_1(-x)}-\mathbf {K(x,y)}& -\mathbf {P_1(-y)} \mathbf {Q_0(y) }\end{bmatrix}
\ee
where $\mathbf{P_0(x)}=[p_{i-1}(x_j)]_{1\leq i,j\leq N}$, $\mathbf{P_1(y)}=[p_{i-1}^{(1)}(y_j)]_{1\leq i,j\leq N}$ and, after exchanging $p_i$s with $q_i$s, we define the 
remaining symbols 
(see equations \eqref{eq:ps} and \eqref{eq:qs} for definitions) accordingly.  
Moreover, $\mathbf{K(x,y)}=[K(x_i,y_j]_{1\leq i,j\leq N}$.  With this notation in place 
it is  clear that $\mathcal{K}$ admits the following (Bruhat) decomposition: 
\be
\mathcal{K}=\begin{bmatrix} \mathbf {P_0(x)}^T & 0\\
\mathbf {-P_1(-y)}^T & I \end{bmatrix}\begin{bmatrix} 0& 
I\\
I& 0\end{bmatrix}\begin{bmatrix} -\mathbf{K(x,y)}& 0\\
-\mathbf {Q_1(-x)}& \mathbf {Q_0(y) }\end{bmatrix}.  
\ee
Thus $\det \mathcal{K}=\det\mathbf{P_0(x)^T}\det\mathbf{K(x,y)}\det\mathbf{Q_0(y)}=
\det[K_N(x_i,y_j)]\det[K (x_i,y_j)]$.  
{\bf Q.E.D.}\par \vskip 5pt
A comparison of the  definitions (\ref{Hkernels}) with the entries of the Christoffel--Darboux identities  of  Prop. \ref{propCDI}  shows that they are intimately related, in fact
\be
H_{\mu,\nu} (x,y) = {(-)^{\mu+\nu}}  \sum_{j=0}^{N-1} q_j^{(\mu)}\big( (-)^\mu x\big) p_j^{(\nu)}\big ( (-)^\nu y\big)  - \frac {\delta_{\mu,1}\delta_{\nu,1}}{x+y}
\ee
and we recognize that (up to some signs) these are precisely the entries of the $\mathbb H_n$ kernel of Prop. \ref{Hkernel}:
\be
H_{\mu,\nu} (x,y) =(-)^{\mu+\nu} \mathbb H_{n,\mu,\nu}\le( (-)^\mu x ,  (-)^\nu y\ri) 
\ee

More explicitly we have  
\bp
The kernels of the correlation functions are given in terms of the solution of the RHP in Prop. \ref{RHP2} as follows 
\be
\begin{array}{cll}
H_{00}(x,y)&  = \mathbb H_{N,00}(x,y) &\ds = \frac {\le[ \wh \Gamma^{-1}(-y) \wh \Gamma(x)\ri]_{3,1} }{x+y} \\[10pt]
H_{01}(y',y) & = -\mathbb H_{N,01}(-y',y) &\ds =\frac {\le[ \wh \Gamma^{-1}(-y) \wh \Gamma(-y')\ri]_{3,2} }{y'-y}\\[10pt]
H_{10}(x,x')  &= -\mathbb H_{N,10}(x,-x')&\ds = \frac {\le[ \wh \Gamma^{-1}(x') \wh \Gamma(x)\ri]_{2,1} }{x'- x} \\[10pt]
H_{11}(y,x)  &= \mathbb H_{N,11}(-y,-x) &\ds = \frac {\le[ \wh \Gamma^{-1}(x) \wh \Gamma(-y)\ri]_{2,2} }{x+y}
\end{array}
\ee
\ep

The potential relevance of these formul\ae\  is that it allows one to compute the large $N$  asymptotic behaviour of the correlation functions in terms of the asymptotic behaviour of only  $3$ consecutive biorthogonal polynomials and auxiliary  functions associated with them  that enter in the Riemann--Hilbert formulation of Props. \ref{RHP1}, \ref{RHP2}.

The steepest--descent analysis of these problems which will appear in a forthcoming paper.

We expect that,
after a complete description of the asymptotics of the BOPs is obtained, these formul\ae\ can be used to address the issue of universality for this matrix model via the  Riemann--Hilbert approach.

\br Note also that the formula above is not symmetric inasmuch as the BOPs $\p, \q$ play different roles in the Christoffell-Darboux theorem; we could however rewrite the theorem as
\bea
 \sum_{j=0}^{N-1} p_j(x) q _j(y) = \frac {\vec {\p}_{_0,N}^t (x) \mathbb B_N (-x) {\vec {\check  \q}}_{_0,N}(y)}{x+y}\ .
\eea
where $\mathbb B(z)$  was defined in (\ref{CDIkernels}). The auxiliary vector $
\check \q_{_0,N}$ enters in a similar Riemann--Hilbert formulation with the r\^oles of the densities $\alpha$ and $\beta$ interchanged.
\er
\subsection {A multi--matrix model}
\label{Chain}
It is possible to extend the model (\ref{mumeas}) to a chain of matrices with the nearest neighbor interaction 
\be
\frac 1{\det(M_{i} + M_{i+1})^{N+h_i}}
\ee
The ``strength'' of the interaction (i.e. $h_j$'s) may depend on a site along the chain.
Specifically, consider the space $\left (\mathcal H_N^+\right )^{\times R}$
and  a collection of $R$ positive densities $\alpha_1,\dots, \alpha_R$ on $\R_+$.
Define the finite mass measure 
\be
\d \mu(M_1,\dots, M_R) =  \prod_{\ell=1}^{R-1} \frac 1{\det(M_\ell + M_{\ell+1})^{N+ h_\ell}} \prod_{\ell=1}^R  \alpha_\ell(M_\ell) \d M_\ell 
\ee
Using the Harnad--Orlov formula one 
we obtain, up to normalization constant,  the following measure which we denote by the same symbol
\be
\d\mu(X_1,\dots, X_R) = \prod_{\ell=1}^R\alpha_\ell(X_\ell)\Delta(X_\ell)^2  \d X_\ell 
\prod_{\ell=1}^{R-1} \frac {\det \le[ (x_{\ell, i} + x_{\ell+1, j})^{-1-h_\ell}\ri]_{1\leq i,j\leq N}}{\Delta(X_\ell)\Delta(X_{\ell+1})}
\ee

Following the same steps that led to the expression (\ref{reducedmeasure}) as positive  density for $(S_N)^2$--invariant observables,  we obtain the following reduced measure on the spectra (up to a normalization depending only on $h_\ell$'s but not on the measures)
\bea
\d\wt \nu(X_1,\dots, X_R) = \Delta(X_1)\Delta(X_R)\prod_{\ell =1}^{R-1} \det (X_\ell +X_{\ell +1})^{-1-h_\ell } \prod_{\ell =1}^R\alpha_\ell (X_\ell ) \d X_\ell  
\eea
In the case where all interactions are the same $h_\ell =0$ we have 
\bea
\d\wt \nu(X_1,\dots, X_R) =\Delta(X_1)\Delta(X_R) \prod_{\ell =1}^{R-1}\frac 1{ \det (X_\ell +X_{\ell +1})} \prod_{\ell =1}^R\alpha_\ell (X_\ell ) \d X_\ell  
\eea
which can be seen as a generalization of (\ref{reducedmeasure}). 
This model too can be treated with the aid of biorthogonal polynomials that will now satisfy a $R+3$ recurrence relation; note that the length of the recurrence relations does not depend on the ``potentials'', or densities,  $\alpha_j$. This  is in sharp contrast with the usual multi-matrix model with interaction ${\rm e}^{c\tr(M_jM_{j+1})}$ \cite{EynardMehta, BEH_dualityCMP}. To characterize these biorthogonal polynomials, an  $(R+2)\times(R+2)$ Riemann--Hilbert problem can be set up and the strong asymptotics can be dealt with, but the complexity of this problem  definitely warrants a separate paper.
\section{Diagrammatic expansion}
\label{Diagram}
In parallel with the $\frac 1 {N^2}$\green{--} expansion for the Hermitean matrix model and the IZHC two-matrix model, we would like to sketch the similar formal expansion of the model  in terms of colored ribbon graphs.
The weights $\alpha$ and $\beta$ entering the definition (\ref{mumeas}) are assumed to be of the form $\alpha(M_1) = {\rm e}^{ -N\tr (U (M_1))},\ \beta(M_2) = {\rm e}^{ -N\tr (V (M_1))}$ . 
We perform a shift and a rescaling of  the matrices so that $\det(M_1 + M_2) \mapsto \det( \1 - \zeta M_1 -\eta  M_2)$.
Of course the values of $\zeta$ and $\eta$ are suitably restricted to a neighborhood of the origin: however this restriction is irrelevant since the manipulations below are in the sense of formal power series.

The procedure amounts to  a perturbative  Taylor--expansion  around a Gaussian integral. In other words we will be considering a partition function in the form  
\be
\mathcal Z_N := \iint \d M_1 \d M_2 {\rm e}^{ -\frac N 2 \tr\le({M_1}^2 +  {M_2}^2 \ri)   + N\sum_{\ell=1}^\infty  \tr (\zeta M_1 +\eta  M_2)^\ell -N  \tr U_p(M_1) -N\tr V_p(M_2)}. 
\ee
where $U_p, V_p$ are the perturbations of the Gaussian (including a quadratic term as well)  which, for convenience, we parametrize as the following formal series
\be
U_p(x) :=- \sum_{j=1}^\infty \frac{u_j-\zeta^j }j  x^j\ ,\ V_p(y) :=- \sum_{j=1}^\infty \frac{v_j-\eta^j}j y^j
\ee
We thus have
\be
\mathcal Z_N = \le< 
\exp N \tr  \le({\sum_{j=1}^\infty\frac {u_j-\zeta^j}j  {M_1}^j   +\sum_{j=1}^\infty\frac {v_j-\eta^j}j   { M_2}^j} + \sum_{\ell=1}^{\infty} \frac{1}\ell \le( (\zeta M_1 + \eta M_2)^\ell\ri)\ri)
\ri>
\ee
where the average is taken w.r.t. the underlying ({\em uncoupled!})  Gaussian measure
\be
\frac 1{\mathcal Z^{(0)}_N} \d M_1 \d M_2\exp \le[-\frac N 2 \tr\le( M_1^2 +  M_2^2 \ri)  \ri]\ ,\ \ {\mathcal Z^{(0)}_N}:=  \int\d M_1 \d M_2\exp \le[-\frac N 2 \tr\le( M_1^2 +  M_2^2 \ri)  \ri]
\ee
Note that, due to the shifts in $u_j, v_j$, the $\ell=1$ term in the third sum above cancels exactly  against the shifts in the first two sums; similarly, only the term $\frac {\zeta\eta}2 \tr (M_1M_2) $ remains in the quadratic part:
\bea
\mathcal Z_N = \le< 
\exp N \tr \le( u_1  {M_1} +  \frac {u_2}2 {M_1}^2    +v_1 { M_2}+ \frac{v_2}2{M_2}^2  + \zeta\eta M_1M_2 + \dots \ri)
\ri>
\eea

Using Wick's theorem for the evaluation of Gaussian integrals and the frequently used combinatorial interpretation  (see \cite{diFrancesco} for an excellent introduction) one sees that the partition function is the sum over all possible {\bf bi-colored} ribbon graphs respecting rules which we now specify

\begin{itemize}
\item There are two colors for vertices/edges (say, red/blue); vertices are distinguished in monochromatic and bichromatic.
\item Each monochromatic vertex of valency $j$ enters with a weight $N u_j/j$ (blue) or $N v_j/j$ (red).
\item Each edge (red or blue) enters with a weight $\frac 1 N$.
\item Each bichromatic vertex of valency $\ell$ enters with a weight $\frac {\zeta^k\eta^{\ell-k}}{N\ell} $ where $1\leq k \leq \ell-1$ is the number of blue half-edges and $\ell-k$ the number of red ones and appears in with a multiplicity $\le({\ell\atop k }\ri)$ corresponding to all possible arrangements of $k$ blue legs amongst $\ell$, up to cyclic reordering. In particular there is only one bichromatic bivalent  vertex  (up to automorphism) which enters with weight $\zeta\eta/N$.
In general, since each such vertex corresponds to a trace of the form $\tr ({M_1}^{a_1} {M_2}^{b_1} {M_1}^{a_2}\dots)$ and in view of the cyclicity of the trace, there are  precisely $\ell$ equivalent vertices obtained by cyclically permuting the matrices in the sum, which corresponds diagrammatically to a rotation of the colors of the legs of the  vertex. Hence there are in fact $\frac 1 \ell \le({\ell \atop k} \ri) = \frac {(\ell-1)!}{(\ell-k)^! k!}$ inequivalent bicolored $\ell$--valent vertices in each diagram contributing with a weight $\zeta^{k} \eta^{\ell - k} /N$.
\item Each connected Feynman  diagram $\Gamma$ constributing to the perturbative sum has a power $N^{F-E+V} = N^{2-2g}$ where $g = g(\Gamma)$ is  the genus of the  surface over which the graph can be drawn.  
\end{itemize}

\begin{wrapfigure}{r}{0.3\textwidth}
\resizebox{6cm}{!}{\includegraphics{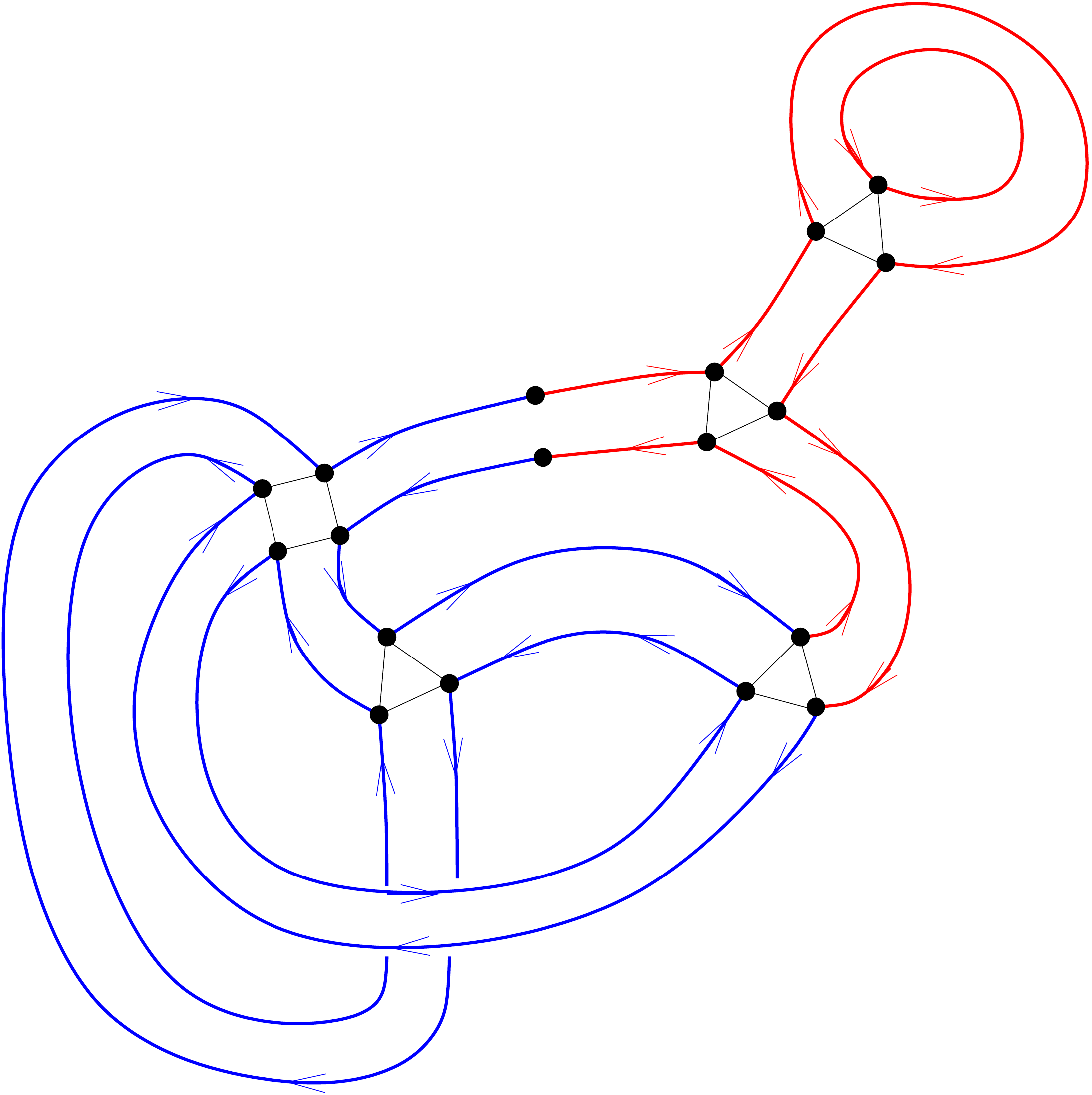}}
\caption {\em An example of a connected diagram contributing to the partition function.}
\end{wrapfigure}

Summing up over all possible labeling of the bicolored fat-graphs $\Gamma$ leaves a factor $|Aut(\Gamma)|$ in the partition function (see \cite{diFrancesco}) and the result is hence
\bea
\ln \mathcal Z_N =\hspace{-0.8cm} \sum_{\hbox{ \small \parbox{2cm}{ Connected bichromatic fatgraphs $\Gamma$}}} 
\hspace{-0.5cm}\frac {N^{2-2g_\Gamma}}{|Aut(\Gamma)|} \prod_{j=1}^\infty   u_j^{n_j} v_j^{m_j}\prod_{\ell=2}^\infty\prod_{k=1}^{\ell-1} \zeta^{(\ell-k)r_k ^\ell} \eta^{k r_k^\ell}
\eea
where $n_j = n_j(\Gamma)$ is the number of blue vertices with valency $j$, $m_j = m_j(\Gamma)$ the number of $j$--valent red  vertices and  $r_k^\ell = r_k^\ell(\Gamma)$ is the number of  $\ell$--valent bichromatic vertices with $k$ red (and hence $\ell-k$ blue) legs.

\section{Large $N$ behaviour}
\label{seclargen}

Consider two densities  $\alpha(x), \beta(y)$ of the form 
\be
\alpha(x) = \alpha_\hbar (x) = {\rm e}^{-\frac 1\hbar V(x)}\ ,\ \ \ \beta(y) = \beta_\hbar (y)  = {\rm e}^{-\frac 1\hbar U(y)}\ ,\qquad 
\hbar = \frac T N, \ T>0\ .
\ee
Here we have introduced a dependence on the small parameter $\hbar$ on the measures (but we will not emphasize this dependence in the notation). 
From the experience amassed in the literature on the ordinary orthogonal polynomials we start by considering the heuristic ``saddle--point'' for the partition function $\mathcal Z_N$, namely the total mass of the reduced measure (\ref{reducedmeasure}). The heuristics calls for a scaling approach where we send the size of the matrices $N$ to infinity and the scaling parameter $\hbar$ to zero as $\mathcal O(1/N)$, namely that we vary the densities $\alpha, \beta$ by raising them to the power $\frac 1 \hbar$. 
We thus have
\be
\mathcal Z_N  =  \int \Delta(X)\Delta(Y) \det \le[\frac 1 {x_i+y_j} \ri]_{i,j\leq N} \alpha(X)   \beta(Y)  \d X \d Y\ .
\ee
Using \ref{Cauchybello} we have 
\be
\mathcal Z_N \propto \int \frac{\Delta(X)^2\Delta(Y)^2}{\prod_{i,j}(x_i+y_j)} \exp\le[-\frac 1 \hbar \sum_{i=1}^N U(x_i) + V(y_j) \ri]\d X\d Y
\ee
Taking  $\-\frac {1} {N^2}$ times the logarithm of the integrand we have the expression 
\bea
S(X,Y) &\&:= \frac 1 {TN} \sum_{j=1}^N U(x_j)+ V(y_j) -  \frac {1} {N^2} \sum_{j\neq k} \ln|x_j-x_k| - \frac {1}{N^2} \sum _{j\neq k} \ln |y_j-y_k|  + \frac {1} {N^2} \sum_{j,k} \ln |x_j + y_k|   =\cr
&\& = S_U(X) + S_V(Y) + J(X,Y)\label{Action}.
\eea

It is convenient --in order to deal with a more standard potential-theoretic problem-- to map $Y\to -Y$ and define  $V^\star(y) = V(-y)$ so that  we can rewrite the action as 
\be
S(X,Y)  = S_U(X) + S_{V^\star}(Y) + \frac {1}{N^2} \sum_{j,k} \ln |x_j-y_k|
\ee
which describes the energy of a gas of $2N$ particles of charge $+1$ (for the $x_j$'s) and $-1$ (for the $y_j$'s) 
separated by an impenetrable, electrically neutral, partition confining the positively charged particles to the positive real axis under the potential $U$ and the negatively charged  particles to the negative axis under the potential $V^\star$. 


The usual argument is that the configuration of the minimum contribute the most to the integral and  --under suitable assumptions of regularity for the potential-- one wants to show that the sequence of these minima configurations tends in-measure to some  probability distributions.

We define
\be
\rho =\frac 1 N \sum_{k=1}^N \delta_{x_k}\ ,\qquad \mu = \frac 1 N \sum_{j=1}^N \delta_{y_j}
\ee
and rewrite the action as 
\bea
S[\rho,\mu] := \frac 1 T \int_{R_+}U(x) \rho(x)\d x-  \int_{\R_+} \int_{\R_+} \rho(x)\rho(x') \ln |x-x'| \d x\d x' +\cr  + \frac 1 T \int_{\R_-} V^\star(y) \mu(y) \d y -  \int_{\R_-} \int_{\R_-} \mu(y)\mu(y') \ln |y-y'| \d y\d y'  +\cr   \int_{\R_+} \int_{\R_-} \rho(x)\mu(y) \ln |x-y| \d x\d y
\eea
\subsection{Continuum version: cubic spectral curve and solution of the potential problem}
We immediately rephrase the above minimization problem in a continuum version. In order to emphasize the symmetry of the problem it is convenient to denote $U(x)$ by $V_1(x)$ and $V^\star(y) = V(-y)$ by $ V_2(y)$ and denote the corresponding equilibrium densities  by $\rho_1(x),\ \rho_2(y)$. 

The main point of this section is Thm. \ref{speccurve} which states that the resolvents (Markov functions)  of the equilibrium distributions are solutions of a cubic equation that defines a trigonal curve\footnote{Namely a curve of the general form $w^3+A w^2+ B w +C=0$, with $A,B,C$ smooth functions of the spectral parameter $z$.}; this result is the analogue of the better-known result for the equilibrium measure appearing in the one--matrix model (\cite{McLaughlinDeiftKriecherbauer} and references therein). The derivation that we present here is of a formal heuristic nature inasmuch as we discount several important issues about the regularity of the equilibrium measures. However this sort of manipulations is quite common and they can be obtained also from the {\em loop equations} as done in \cite{EynardOn} for the $O(n)$ model. 

Rewriting the functional in the new notation for the potentials we find
\bea
S[\rho_1,\rho_2] := \frac 1 T \int_{R_+} V_1(x) \rho_1(x)\d x-  \int_{\R_+} \int_{\R_+} \rho_1(x)\rho_1(x') \ln |x-x'| \d x\d x' +\cr  + \frac 1   T\int_{\R_-} V_2(y) \rho_2(y) \d y -  \int_{\R_-} \int_{\R_-} \rho_2(y)\rho_2(y') \ln |y-y'| \d y\d y'  +\cr   + \int_{\R_+} \int_{\R_-} \rho_1(x)\rho_2(y) \ln |x-y| \d x\d y
\eea
We will be studying the measures $\rho_1,\rho_2$  that minimize the above functional; here we will {\bf assume} their regularity (which shall be proved in a separate paper). Also, the potentials $V_1,V_2$ will  be  assumed {\bf real analytic}.

More precisely, we make the following assumption on the nature of the potentials
\begin{assumption}
\label{assumptionV}
The two potentials $V_1(x), V_2(y)$ are restrictions to the positive/negative axis of two real--analytic functions such that
\begin{itemize}
\item $\lim_{x\to +\infty} \frac {V_1(x)}{\ln x} = +\infty$, and similarly $\lim_{y\to -\infty} \frac {V_2(y)}{\ln |y|} = +\infty$;
\item the derivatives  $V_j'(x)$ have at most finitely many poles in a strip of finite width around $\R$; 
\item finally,  $V_1(x) > C_1 |\ln (x)|$ for some constant $C_1>0$ and $x>0$ and $V_2(y)> C_2 | \ln |y|\,|$ for $y<0$ and some constant $C_2>0 $.
\end{itemize}
\end{assumption}
It will be shown in a  separate paper that this assumption is sufficient to guarantee that the equilibrium densities exist, have compact support, are regular and their supports do not include the origin. 
In order to enforce the normalization of $\rho_1,\rho_2$ --as customary-- we will introduce in the action two suitable Lagrange multipliers
\be
\wh S : = \wh S[\gamma_1,\gamma_2,\rho_1,\rho_2]:= S[\rho_1,\rho_2] + \gamma_+ \le(\int \rho_1 \d x -1\ri)  + \gamma_- \le(\int \rho_2 \d x -1\ri) 
\ee

The vanishing of the first variation of the functional $\wh S$ yields the equations
\bea
&& \frac {V_1(z)}T - 2 \int_{\R_+} \rho_1 \ln |z-x|\d x  + \int_{\R_-} \rho_2 \ln|z-x|\d x  = \gamma_+ \ ,\  z\in\,Supp(\rho_1) \cr
&& \frac {V_2(z)}T - 2 \int_{\R_-}  \rho_2 \ln |z-x|\d x  + \int_{\R_+} \rho_1 \ln|z-x|\d x  = \gamma_- \ ,\  z\in\,Supp(\rho_2) 
\eea
Differentiating w.r.t. $z$ yields 
\bea
\frac 1{T}V_1'(z)  -2  P.V. \int \rho_1 \frac1{z-x} \d x  +  \int \rho_{2} \frac 1 {z-x}\d x  = 0\ ,\  z\in\,Supp(\rho_1)\cr
\frac 1{T}V_2'(z)  -2  P.V. \int \rho_2\frac1{z-x} \d x  +  \int \rho_{1} \frac 1 {z-x}\d x  = 0\ ,\  z\in\,Supp(\rho_2)
\eea
where $P.V.\int$ indicates the Cauchy principal value. These equations are best written in terms of  the {\bf resolvents} (or {\bf Weyl functions})
\be
W_i(z):= \int \frac {\rho_i(x) \d x}{z-x},\ \qquad z\in \C \setminus Supp(\rho_i)\ .
\ee
Indeed, with the help of the Sokhotskyi--Plemelj formula,  the equations above  take a simpler form 
\bea
\le\{
\begin{array}{rl}
W_{1,+}(z) + W_{1,-}(z) & =\frac 1T V_1'(z)  + W_{2} (z) \ , \ z\in Supp(\rho_1)\cr
W_{2,+}(z) + W_{2,-}(z) & =\frac 1 T  V_2'(z)  + W_{1} (z) \ , \ z\in Supp(\rho_2)
\end{array}\ri.
\\
W_{i,+} - W_{i,-} &\& = -  2i\pi \rho_i
\label{Wjumps}
\eea
Note that in the RHS of the system of  equations,  the resolvent of the other measure is evaluated at a regular point due to the fact that the two supports are disjoint.

We define, partly motivated by 
\cite{EynardZinn}, the shifted resolvents

\be
Y_1:= -W_1 + \frac {2V_1'+V_2'}{3T}\ ,\qquad
Y_2:= W_2 - \frac {V_1'+2V_2'}{3T}\label{whys}.  
\ee
 Observe that, in view of Assumption (\ref{assumptionV}) , 
$\{Y_1, Y_2\}$ have the same analytic structure as $\{W_1, W_2\}$, 
while the equations describing the jumps simplify to: 
\bea
\le\{
\begin{array}{rl}
Y_{1,+} + Y_{1,-} & = - Y_2 (z) \ , \ z\in Supp(\rho_1)\cr
Y_{2,+} + Y_{2,-} & =  -Y_1 (z) \ , \ z\in Supp(\rho_2)
\end{array}\ri.
\label{jump1}
\eea 
\be
 Y_{1,+} - Y_{1,-} = 2i\pi \rho_1\ ,\qquad
Y_{2,+} - Y_{2,-}  = -2i\pi \rho_2
\label{jump2}
\ee
Using these equations one obtains by direct inspection that 
\be
R(x):= {Y_1}^2 + {Y_2}^2 +Y_1Y_2  \label{Rcoeff}
\ee
has no jumps on either supports.
Multiplying eq. (\ref{Rcoeff}) on both sides by $Y_1-Y_2$  one obtains
\be
R(z) (Y_1-Y_2) = {Y_1}^3 - {Y_2}^3
\ee
which can be rewritten as 
\be
{Y_1}^3 - R(z) Y_1 = {Y_2}^3 - R(z) Y_2:=D(z) \label{Dcoeff}\ .
\ee
The first expression may  {\em a priori} have at most jumps across the support of $\rho_1$, while the second may have jumps only across the support of $\rho_2$: since the two supports are disjoint we conclude that $D(x)$ is a regular function on the real axis.
\br
\begin{figure}
\resizebox{0.8\textwidth}{!}{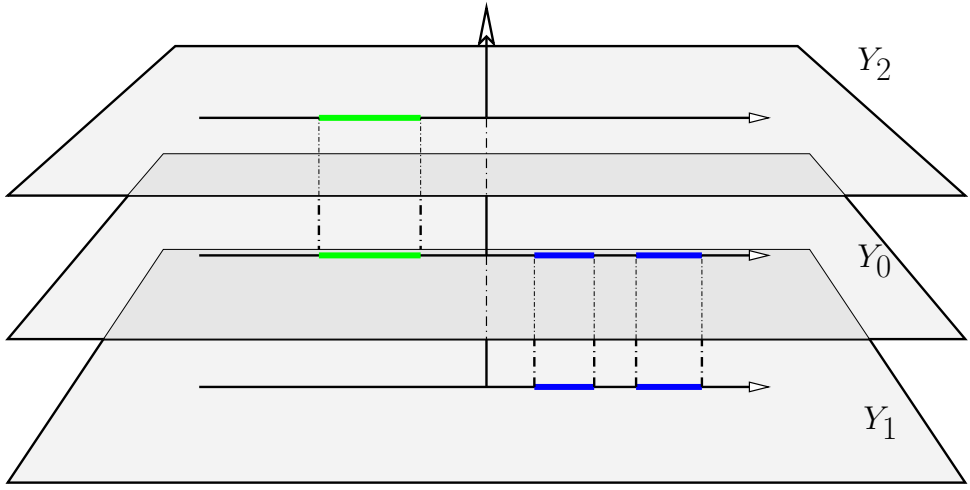}
\caption{A pictorial example of the three sheets of the Riemann surface of $Y(x)$. In blue the support of $\rho_1$ (right) and in green that of $\rho_2$ (left).}
\label{Hurwitz}
\end{figure}
If we introduce the function $Y_0:= -Y_1-Y_2$ the jump relations (\ref{jump1}, \ref{jump2}) can be rewritten (after a few straightforward manipulations) as 
\bea
\begin{array}{cc}
Y_{0,\pm}(z) = Y_{1,\mp}(z) & z\in Supp(\rho_1)\\
Y_{0,\pm}(z)=  Y_{2,\mp}(z) & z\in Supp(\rho_2)\ .
\end{array}
\eea
This implies that we can think of the  $Y_{j}(z)$'s as the three branches (three {\em sheets}) of a Riemann surface realized as a triple cover of the $x$--plane branched at the endpoints of the spectral bands. The Riemann surface is depicted in Fig. \ref{Hurwitz}. Thm. \ref{speccurve}  will be proving more formally this statement, realizing $Y_0,Y_1, Y_2$ as branches of a cubic equation (which the reader can already guess from (\ref{Dcoeff})).
\er
\bt
\label{speccurve}
Under Assumption (\ref{assumptionV}),   the two shifted resolvents satisfy the same {\bf cubic} equation in the form 
\be
E(y,z) := y^3 - R(z) y - D(z) = 0\ .\label{algcurve}
\ee

 The coefficients $R(x), D(x)$ are related to the equilibrium measures as follows: 
\bea
R(z) &\& = \frac {(V_1')^2 + (V_2')^2 + V_1' V_2'}{3T^2} -R_1(z) - R_2(z)\cr
&& R_i(z):= \frac 1 T \int \frac {V_i'(z)-V_i'(x)}{z-x}\rho_i(x)  \d x = \frac 1 T \nabla_i V_i' \\
D(x) &\&  = U_0U_1U_2   + U_2 R_1 + U_1 R_2 
+\frac 1 T \nabla_{12} V_1' - \frac 1 T \nabla_{21} V_2' \\
\eea
where the functions $U_0,U_1,U_2$ and the  operators $\nabla_1,\nabla_2, \nabla_{12},\nabla_{21}$ have been defined as 
\bea
&& 
U_1:= \frac {2V_1'+V_2'}{3T}\ ,\ 
U_2:= -\frac {2V_2'+V_1'}{3T}\ ,\ 
U_0:= \frac {V_2'-V_1'}{3T}\label{Ujs}
\\
&& 
\nabla_j f(z):= \int \frac {f(z)-f(x)}{z-x} \d\rho_j(x)\ ,\cr 
&&
\nabla_{12} f(z):= \iint \frac {f(z)-f(x)}{z-x} \frac {\d\rho_1(x)\d\rho_2(y)}{x-y}\ ,\\ 
&&
\nabla_{21} f(z):= \iint \frac {f(z)-f(x)}{z-y} \frac {\d\rho_2(y)\d\rho_1(x)}{y-x}
\eea

Furthermore the support of $\rho_1,\rho_2$ must consist of {\bf finite union of finite disjoint intervals}.
\et

{\bf Proof}.
The statement about the support of the measures follows from the argument used in \cite{McLaughlinDeiftKriecherbauer} that that we sketch here.
By results of  Chapter 13 of \cite{SaffTotik}),  Assumption \ref{assumptionV} implies that the supports of the equilibrium measures are compact. Moreover, it follows that the branchpoints (and branchcuts) of the functions $Y_{1,2,0}$ coincide with the supports of $\rho_1,\rho_2$. Since they solve a cubic algebraic equation, the branchpoints are determined as zeroes of the discriminant $\Delta:= 4R^3 - 27 D^2$ of (\ref{algcurve}).

As suggested by the previous discussion, even without an explicit expression for $R,D$, it follows from Morera's theorem that $R, D$ are also real-analytic and so is $\Delta$. Thus $\Delta$ cannot have infinitely many zeroes in a compact domain and so the number of endpoints of the supports is a-priori finite.

The only part that is left to be proven are the formul\ae\ for $R, D$.

Computing the jump of ${W_j}^2$ from  eqs. (\ref {Wjumps}) one obtains 
\be
(W_1)^2_+- (W_1)^2_- = -2i\pi \rho_1 \le( \frac 1  T V_1' + W_2\ri)
\ee
 and a similar equation for $W_2$. This implies that 
\bea
{W_1}^2(z) &\& = \int \frac {\rho_1(x) \le(V_1'(x) /T+ W_2(x) \ri)}{z-x}\d x  =\cr 
&\&= 
\frac 1  T V_1'(z)W_1(z) -
\underbrace{ \int\frac {\rho_1(x) \le(V_1'(z)-V_1'(x)\ri)}{T(z-x)}\d x}_{=: R_1(z)}  + \iint \frac {\rho_1(x) \rho_2(y)\d x \d y}{(z-x)(x-y)} 
\eea
Note that $R_1(z)$ is regular on the support of $\rho_1$ since $V_1'$ is.
Hence we have 
\bea
{W_1}^2(z)  =\frac 1 T  V_1'(z)W_1(z) - R_1(z) +\overbrace{ \iint \frac {\rho_1(x) \rho_2(y)\d x \d y}{(z-x)(x-y)}}^{=: W_{12}}
\label{W1square}\\ 
{W_2}^2(z)  =\frac 1  T  V_2'(z)W_2(z) - R_2(z) + \underbrace{\iint \frac {\rho_1(x) \rho_2(y)\d x \d y}{(z-y)(y-x)}}_{=:W_{21}}
\label{W2square} \ .
\eea
Adding the two together and using the identity 
\be
\frac 1{(z-x)(x-y)} + \frac 1{(z-y)(y-x)} = \frac 1{(z-x)(z-y)}
\ee
we obtain 
\be
{W_1}^2 + {W_2}^2  = \frac 1   T V_1' W_1 +\frac 1  T  V_2'W_2 - R_1 - R_2 + W_1 W_2\ ,
\ee
which is precisely eq. (\ref{Rcoeff}) when rewritten 
with $W_1, W_2$ replaced by their expressions in terms of the $Y_1, Y_2$ variables resulting from (\ref{whys}).

The second formula for $D(x)$ can be obtained by noticing that 
$D(x) = Y_0Y_1Y_2$.
Explicitly, this reads  (after rearranging the terms and using the definitions for the shifted resolvents (\ref{whys}) and the $U_j$'s (\ref{Ujs}) 
\bea
D(x) = U_0U_1U_2  -2U_0 W_1W_2 - W_1^2W_2 + W_1 W_2^2  - U_2\le( W_1^2-\frac{V_1'}T W_1\ri)- U_1 \le(W_2^2  - \frac{V_2'}T W_2\ri )
\eea

Substituting in the last two terms the expressions (\ref{W1square}, \ref{W2square}) we obtain 
\bea
D(x) =  U_0U_1U_2  -2U_0 W_1W_2 - W_1^2W_2 + W_1 W_2^2  + U_2 R_1 + U_1 R_2 - U_2 W_{12}
  - U_1W_{21}
  \eea
Using  $ W_1 W_2= W_{12}+W_{21} $ (\ref{W1square}, \ref{W2square}) we obtain 
\be
D(x) =  U_0U_1U_2   - W_1^2W_2 + W_1 W_2^2  + U_2 R_1 + U_1 R_2 
+\frac{V_1'}T W_{12} - \frac {V_2'}T W_{21}
\ee
Now we can rewrite 
\bea
V_1'W_{12} = \nabla_{12} (V_1') + \iint \frac {V_1'(x)\rho_1(x)\rho_2(y)\d x \d y}{(z-x)(x-y)} \\
V_2'W_{21} = \nabla_{21} (V_2') + \iint \frac {V_2'(y)\rho_1(x)\rho_2(y)\d x \d y}{(z-y)(y-x)} 
\eea
so that we have obtained
\bea
D(x)&\&  =  U_0U_1U_2   + U_2 R_1 + U_1 R_2 
+\frac 1 T \nabla_{12} V_1' - \frac 1 T \nabla_{21} V_2' + \mathfrak R\\
&\& \mathfrak R := \frac 1 T    \iint\le( \frac {V_1'(x)}{z-x} + \frac {V_2'(y)}{z-y}\ri) \frac{\rho_1(x)\rho_2(y)\d x \d y}{(x-y)}  
 - W_1^2W_2 + W_1 W_2^2 
\eea
We want to show that $\mathfrak R\equiv 0$: indeed it is clear that the only part of $D(x)$ which may have jump-discontinuities is $\mathfrak R$, but we know that $D(x)$ has no such discontinuities.

Hence $\mathfrak R$ has no discontinuities; on the other hand it is clear from its definition that $\mathfrak R$ cannot have any other singularities and hence it is an entire function. Inspection shows that $\mathfrak R(z) \to 0$ as $z\to \infty$ and hence $\mathfrak R$ must be identically vanishing by Liouville's theorem. This concludes the proof of the theorem. {\bf Q.E.D.}\par

 The presence of a ``spectral curve'' will be one of the crucial ingredients for the large--degree asymptotic analysis of the  biorthogonal polynomials using the Riemann--Hilbert formulation given in \cite{Paper1} and the Deift--Zhou nonlinear steepest descent method. Indeed  we will show in a separate publication that the OPs are modeled by (spinorial) Baker--Akhiezer vectors (similarly to \cite{BertoMo}) that naturally live on the three-sheeted covering specified by  (\ref{algcurve}).

\vskip 5pt

 \appendix
 \renewcommand{\theequation}{\Alph{section}-\arabic{equation}}

 \section{A rectangular mixed  $3$--matrix model with ghost fields}
\label{rectangular}
The model (\ref{muh})  (for any value of $h$) can be obtained from the following matrix models
\subsection{Integer $h<N$}
Consider the standard Lebesgue measure  on the space $Mat((N-h)\times N,\C)$ of
complex $(N-h)\times N$ matrices, viewed as a linear space: if $A\in Mat((N-h)\times N,\C)$ we will
use a shorthand notation
\be
\d A \d A^\dagger := \prod_{i=1}^{N-h} \prod_{j=1}^N \d \Re A_{ij} \d \Im A_{ij}\ 
\ee
for the volume element.
Let, as above, $M_1,M_2 \in \mathcal H_{N}^+$ and consider the following normalizable measure on the space $\mathcal H_N^+\times \mathcal H_N^+ \times Mat(N-h,N,\C)$
\be
\d \wh \mu_\C (M_1,M_2,A):=  \d M_1 \d M_2 \d A \d A^\dagger \alpha(M_1)\beta(M_2) {\rm e}^{- \tr A (M_1+M_2) A^\dagger}
\ee
Since the measure is Gaussian  in $A$  one immediately sees that (up to inessential proportionality constants) 
\bea
\int _{Mat(N-h,N,\C)} \d \wh \mu_\C (M_1,M_2,A) &\& \propto \d \mu_h  (M_1,M_2) \cr
&\& = \d M_1 \d M_2 \frac{\alpha(M_1) \beta(M_2)} {\det(M_1+M_2)^{N-h}}\ , \cr
&\&  \ h \in \{0,1,\dots, N-1\}
\eea
\subsection{Half--integer $h<N$}
One could also consider a similar model where $Mat(N-h,N,\C)$ is replaced by $Mat(2N-2h,N,\R)$ where $h\in \frac 1 2 Z$ and $h<N$  obtaining then 
\bea
\d A&\& := \prod_{i=1}^{2N-2h} \prod_{j=1}^N \d A_{ij} \cr 
\d \wh \mu_\R (M_1,M_2,A)&\& :=  \d M_1 \d M_2 \d A\alpha(M_1)\beta(M_2) {\rm e}^{- \tr A (M_1+M_2) A^t}\\
\int _{Mat(N-h,N,\R)} \d \wh \mu_\R (M_1,M_2,A) &\& \propto \d \mu  (M_1,M_2) = \cr
&\& =\d M_1 \d M_2 \frac{\alpha(M_1) \beta(M_2)} {\det(M_1+M_2)^{N-h}}\ , 
\eea

Of course, if $h$ is actually an integer then this case reduces to the previous one.
\subsection{Integer and half integer $h>N$}
For the sake of  completeness we note that when $h>N$  the determinant in the reduced measure (\ref{muh}) is a positive power in the numerator;  this can be obtained from a Gaussian integral over {\bf Grassmann} variables which can be obtained from ``complex'' anticommuting variables (for $h$ integer) or ``real'' (for $h \in \frac 1 2 \Z$) much along the line of the previous (commuting variable) case.\par \vskip 10pt

Clearly, the plethora of models is endless and each value of $h$ can be studied with the aid of  a different set of biorthogonal polynomial.

We singled out the case $h=0$, since it corresponds to the   Cauchy BOPs that have many features in common with the much better--known orthogonal polynomials and yet seem to have a very rich but still tractable asymptotic theory.

 \section{Relation to the $O(1)$--model}
 \label{Onn}
 The $O(n)$ matrix model \cite{EynardOn} is a multimatrix model for a (positive) Hermitean matrix $M$ and $n$ Hermitean matrices $A_j$ with distribution 
 \be
 \d M \prod_{j=1}^n \d A_j \exp\le(-N \tr \le[V(M) + M\cdot \sum_{j=1}^n {A_j}^2\ri]\ri)\label{Onmeasure}\ .
 \ee
With some manipulations, the integration over the Gaussian variables $A_j$ can be performed and the result written in terms of the eigenvalues $z_j$ of the matrix $M$, yielding a (unnormalized) measure over the space of eigenvalues given by  
\be
\frac {\Delta(Z)^2}{\prod_{i<j} (z_j+z_i)^n}\prod_{j=1}^N \frac{\d z_j{\rm e}^{-N V(z_j)} }{z_j^{\frac n2}} =: \frac {\Delta(Z)^2\alpha(Z)\d Z}{\prod_{i<j} (z_j+z_i)^n}\ ,\ \ \ Z:= diag(z_1,\dots, z_N)\label{O1}
\ee
We wish to show that we can specialize our Cauchy two-matrix-model so that,  in principle, it reduces to a $O(1)$ model.  More precisely, we will show that
 the partition function of the Cauchy two--matrix model is the square of the partition function of the $O(1)$ model for a particular choice of measures (see Prop. \ref{B2}) .

 We recall that from a diagrammatic standpoint the large $N$ expansion of the $O(n)$ model  describes formally a gas of self-avoiding loops of $n$ different colors (hence in our case of a single color) on random surfaces.
 \bp
 \label{B2}
 The partition function of the Cauchy $2$-matrix-model with $N=2k$ and  $\beta(x) = x\alpha(x)$ is the square of the partition function of the $O(1)$-model of the same size. More precisely
 \bea
\le( \frac 1{2^k k! N!}\int_{\R_+^{2k}} \frac {\Delta(Z)^2\alpha(Z)\d Z}{\prod_{i<j} (z_j+z_i)^n}\ri)^2 =\frac 1{N!}
\int_{\R_+^{2k}}\int_{\R_+^{2k}}\frac {\prod_{j=1}^{2k} y_j \Delta(X)\Delta(Y)\a(X)\a(Y) \d X \d Y}{\det (X+Y)}\label{B3}
 \eea
 \ep
 \br
  We conjecture an analogous statement to hold for odd $N$.
 \er
 {\bf Proof.}
 The matrix of moments for this choice of measures reads $I_{ij} = \iint x^{i} y^{j+1} \a(x) \a(y)\frac{\d x \d y}{x+y}$  and the partition function for the corresponding  size--$N$ Cauchy matrix model is by equation \eqref{partminor}  (up to some combinatorial coefficients) the principal minor of this matrix $\det (I)_N$ (we use the subscript $N$ to denote the principal minor of size $N$). Such minor coincides with the RHS of (\ref{B3})
 
 We introduce the new  the skew symmetric  matrix 
 \bea
 M_{ij} = \frac 1 2 \iint x^i y^j (y-x)\frac{\d^2\a}{x+y} 
 \eea
 which is clearly skew--symmetric.  Denoting by $\alpha = [\int x^j \a(x)\d x]^t_{j=0,1,\dots}$ the infinite vector of moments of the measure $\a(x) \d x$, 
 we see that 
 \be
 I = M+ \frac 1 2 \a \a^t
 \ee
  and hence 
 \be
 \det(I_N) = \det (M_N) + \frac 12 \a^t \wt{ M_N}\a
 \ee
 where the tilde denotes the classical adjoint matrix of the principal minor of size $N\times N$ (the transposed of the matrix of cofactors). 

Since  we are dealing with the case $N=2k$ even, the adjoint of a skew--symmetric matrix is skew-symmetric as well and hence the second term has to vanish.  Thus we have
\be
\det (I)_N = \det M_N  = (Pf(M_N))^2
\ee
\def\v{{\bf v}}
%
We see that up to a multiplicative factor $\frac{1}{k! 2^k}$ 
 \begin{equation*}
 \begin{split}
k!2^k \Pf (M_N) = \sum_{\sigma\in S_N} \epsilon(\sigma)\prod_{j=1}^{k} M_{\sigma_{2j-1}\sigma_{2j}} =\\
\int_{\R^N}\int_{\R^N} \prod_{j=1}^{k}\frac{\a(x_j) \a(y_j) \d x_j\d y_j}{x_j+y_j} \det \le[
 \begin{array}{ccccc}
 1 & y_1 &\dots & 1 & y_k\cr
 x_1& y_1^2& \dots & x_{k} & y_k^2\cr
 \vdots &&&&\vdots\cr
 x_1^{N-1} & y_1^{N} & \dots & x_{k}^{N-1} & y_k^{N}
 \end{array}
 \ri]\end{split}
\end{equation*} 
 We denote by $Z$ the vector of size $N$ with components
 \be
 Z = (z_1,z_2,\dots, z_N) = (x_1,y_1,x_2,y_2,\dots, x_k, y_k)\ .
 \ee
 With this notation in place we have 
 \bea 
 k!2^k
 \Pf M_N = \int_{\R^N} \a(Z)\d Z \Delta(Z)  \prod_{j=1}^{k} \frac {y_j} {x_j+y_j} = \cr=
 \int_{\R^N}\underbrace{ \prod_{j<\ell\leq N} \frac 1{z_j + z_\ell}}_{\Gamma(Z)} \underbrace{  \prod_{j=1}^k y_j  \prod_{j\neq \ell\leq k } (x_j + y_\ell) \prod _{j<\ell \leq k} (x_j + x_\ell)(y_j + y_\ell)}_{=: R(Z)}   \a(Z)\Delta(Z) \d Z 
 \eea
 The rational function $\Gamma(Z)$ is invariant under permutations of the variables, whereas $R(Z)$ is not; however $R(Z)$ has the same total degree as  $\Delta(Z)$,  in particular its degree in $y_1$ is $N-1$.
 
 Symmetrizing under the integral sign the variables gives:
 \be
 k!2^k\Pf M_N = \frac 1 {N!} \sum_{\sigma \in S_N} \int \Gamma(Z_\sigma ) \Delta(Z_\sigma) R(Z_\sigma)   \a(Z)\d Z = \frac 1 {N!}\int  \Gamma(Z) \Delta(Z)  \sum_{\sigma \in S_N}\epsilon(\sigma) R(Z_\sigma) \a(Z) \d Z. 
 \ee
 Now note that necessarily 
 \be
  \sum_{\sigma \in S_N}\epsilon(\sigma) R(Z_\sigma)  = \Delta(Z). 
 \ee
 Indeed $R(Z)$ containts the monomial $y_1^{N-1}$ (with the coefficient $1$), and has the same total degree in all variables; since the antisymmetrization must be divisible by $\Delta(Z)$ the assertion follows.
 We thus have 
 \be
 \Pf M_N  = \frac 1{2^k k! N!} \int_{\R^N }\frac{\Delta^2(Z)} {\prod_{j<\ell\leq N} (z_j+ z_\ell)} \a(Z)\d Z.
 \ee

 We see then tha up to a proportionality constant $\Pf M_N$ is the total integral of the measure (\ref{O1}) and thus the proposition is proved.
  {\bf Q.E.D.}\par \vskip 5pt

 The relationship between the two model does not seem to go much further in the sense that there is no direct and simple relationship between the correlation functions of the two models. It seems, however, that some connection should be present and is worth exploring.
 We leave it as an open problem to establish a connection between these two 
 models on the level of the correlation functions.  
%

 %
 \bibliographystyle{plain}
\bibliography{BOP}

 \end{document}